\documentclass[prb,twocolumn,showpacs,preprintnumbers,amsmath,amssymb]{revtex4}
\usepackage{dcolumn}
\usepackage{bm}
\usepackage{graphics}
\usepackage{graphicx}
\usepackage{color}
\usepackage{pstricks,pst-node,pst-text}
\usepackage{times}
\input epsf

\newcommand{\new}[1]{ #1 }
\newcommand{\delete}[1]{}

\newcommand{\Tc}{\ensuremath{T_{\mathrm{C}}}}
\newcommand{\muB}{\ensuremath{\mu_{\mathrm{B}}}}
\newcommand{\vc}[1]{\ensuremath{\boldsymbol{#1}}}
\newcommand{\Mn}{\ensuremath{{\rm Mn}}}
\newcommand{\Ni}{\ensuremath{\rm Ni}}

\newcommand{\alat}{\ensuremath{a_{\rm lat}}}
\newcommand{\Eloc}{\ensuremath{E_{\rm loc}}}

\begin{document}

\title{Exchange interactions and local moment fluctuation
  corrections in ferromagnets at finite temperatures based on non-collinear density-functional calculations}
\author{Marjana Le\v{z}ai\'{c}}\email{M.Lezaic@fz-juelich.de}
\author{Phivos Mavropoulos}\email{Ph.Mavropoulos@fz-juelich.de}
\author{Gustav Bihlmayer}
\author{Stefan Bl\"ugel}

\affiliation{Peter Gr\"unberg Institut and Institute for Advanced Simulation, Forschungszentrum
  J\"ulich and JARA, D-52425 J\"ulich, Germany}

\begin{abstract}

  We explore the derivation of interatomic exchange interactions in
  ferromagnets within density-functional theory (DFT) and the mapping
  of DFT results
  onto a spin Hamiltonian. We delve into the problem of systems
  comprising atoms with strong spontaneous moments together with atoms
  with weak induced moments. All moments are considered as degrees of
  freedom, with the strong moments thermally fluctuating only in angle
  and the weak moments thermally fluctuating in angle and
  magnitude. We argue that a quadratic dependence of the energy on the
  weak local moments magnitude, which is a good approximation in many
  cases, allows for an elimination of the weak-moment degrees of
  freedom from the thermodynamic expressions in favor of a
  renormalization of the Heisenberg interactions among the strong
  moments. We show that the renormalization is valid at all
  temperatures accounting for the thermal fluctuations and resulting
  in temperature-independent renormalized interactions. These are
  shown to be the ones directly derived from total-energy DFT
  calculations by constraining the strong-moment directions, as is
  done e.g.\ in spin-spiral methods. We furthermore prove that within
  this framework the thermodynamics of the weak-moment subsystem, and
  in particular all correlation functions, can be derived as
  polynomials of the correlation functions of the strong-moment
  subsystem with coefficients that depend on the spin susceptibility
  and that can be calculated within DFT. These conclusions are
  rigorous under certain physical assumptions on the measure in the
  magnetic phase space. We implement the scheme in the full-potential
  linearized augmented plane wave (FLAPW) method using the concept of
  spin-spiral states, considering applicable symmetry relations and
  the use of the magnetic force theorem.  Our analytical results are
  corroborated by numerical calculations employing DFT and a Monte
  Carlo method.

\end{abstract}

\pacs{71.70.Gm,75.10.Lp,75.30.Et,75.10.-b}

\maketitle

\section{Introduction}

In recent years there has been increasing activity in the prediction
of high-temperature magnetic properties of solids, especially
regarding critical magnetic transition temperatures. The theoretical
approach is founded in many cases on two assumptions: (i) that the
magnetic excitations of a system can be phenomenologically described
within a classical or quantum Heisenberg model, and (ii) that the
exchange parameters entering the model (i.e., the excitation energies)
can be microscopically derived from total energy results of e.g.
density-functional theory calculations. This ``magnetic multi-scale
modelling'' has proven succesful in many cases, including itinerant
elemental ferromagnets such as Fe and
Co,\cite{MacLaren99,Pajda01,Shallcross05,Lezaic07} localized-moment
systems such as Gd,\cite{Turek03,Khmelevskyi07} magnetic alloys as
NiMnSb or Co$_2$MnSi,\cite{Sasioglu05} or diluted magnetic semiconductors as
(Ga,Mn)As.\cite{Sato04,Bergqvist04} In these systems, the estimated
Curie temperatures \Tc\ are within 10-15\% of the experimental values,
showing that the approach is reliable for practical purposes.

The derivation of magnetic excitation energies from density-functional
calculations requires additional assumptions, since density-functional
theory is, in principle, valid only for the description of the ground
state of many-electron systems. Mainly, one relies on an adiabatic
hypothesis, which conjectures that the slow motion of low-energy
magnetic excitations can be decoupled from the fast motion of
intersite electron hopping, so that the local electronic structure has
time to relax under the constraint that a magnon traverses the
system. Then the magnons are regarded as practically static or
``frozen'' objects, and constrained density functional
theory\cite{Dederichs84} is employed. This adiabatic approximation,
together with the realization that the local moments persist
until above the Curie temperature, constitutes a paradigm which has proven
fruitful. Quite a few theories and methods have been based on it,
aimed at the calculation of thermodynamic quantities, including \Tc,
by harvesting the model parameters from first principles calculations
and working out the thermodynamics within Monte Carlo methods or other
suitable approaches.

Among the first to discuss the adiabatic approximation in connection
to DFT calculations were Gyorffy
and co-workers in the development of a mean-field theory of magnetic
fluctuations,\cite{Gyorffy85} even though the concept was applied
earlier on the level of solutions to many-body model Hamiltonians (see
e.g. Small and Heine\cite{Small84}). Later Antropov et al.\cite{Antropov96}
and Halilov et al.\cite{Halilov98} derived equations for adiabatic
{\it ab-initio} spin dynamics. Further elaboration came from Niu and
collaborators\cite{Niu99} and Gebauer and Baroni,\cite{Gebauer00} who
showed that the Berry curvature enters the adiabatic dynamics
equations; they also demonstrated mathematically how the
Born-Oppenheimer method can be generalized for adiabatic spin dynamics
without the requirement that there should be a large mass difference
defining the time-scale difference.

Based on the adiabatic approximation, a proper parametrization of the
excitation spectrum is called for. In many, but not all, ferromagnetic
systems the magnitude of the local atomic moments is relatively robust
under modest rotations, so that mainly the pair exchange constants
entering a Heisenberg model are required.  From the point of view of
methodology, two approaches are commonly used for the determination of
the pair exchange constants. The first is based on multiple-scattering
theory and Green function methods, frequently in the approximation of
infinitesimal rotations.\cite{Liechtenstein87} It is widely used in
methods where the Green function is
available.\cite{FrotaPessoa00,Pajda01,Sato04,MacLaren99,Katsnelson00,Rusz06,Ruban07,Polesya10,Szunyogh11,Dias11}
\new{A variant of this approach is the use of the
  disorder-local-moment (DLM) reference state\cite{Gyorffy85} as an
  approximation to the magnetic disorder at the Curie temperature,
  from which one obtains exchange parameters\cite{Oguchi83,Oguchi83b}
  in a manner analogous to the method of infinitesimal
  rotations.\cite{Liechtenstein87} Higher-order interactions (e.g. the
  biquadratic, 4-th order, or anisotropic exchange) can be also
  obtained in a systematic way either from the ferromagnetic ground
  state\cite{Udvardi03,Ebert09} or from the DLM
  state.\cite{Szunyogh11} }

The second approach, which we follow in this work, is based on
reciprocal-space calculation of spin-wave excitation energies by
constraining the system to spin spirals of given wavevectors
$\vc{q}$. This requires non-collinear calculations, restricted to the
primitive chemical unit cell (in the absence of spin-orbit coupling)
by virtue of a generalized Bloch theorem.\cite{Sandratskii86} A
subsequent integration in $\vc{q}$-space (in form of a back-Fourier
transformation) yields the pair exchange constants. Although the force
theorem\cite{Andersen} is in principle not necessary here, practically
it is often used in order to avoid the numerical load of a
self-consistent calculation for each vector $\vc{q}$. This second
approach is well-suited for electronic structure methods which are
based on Hamiltonian diagonalization rather than Green functions, and
has been developed, for example, for the augmented spherical wave
(ASW) method\cite{Sandratskii02,Kubler06} or the LMTO
method.\cite{Halilov98}  \new{Concerning the second approach, we
  should also mention that a proper energy fit to a large number of
  constrained-moment-directions has been used to extract model
  parameters, inspired by alloy-theory methods, for example by
  applying the Connolly-Williams theory\cite{Connolly83} to fit a
  number of antiferromagnetic states\cite{Peng91} or in a more refined
  way by applying a systematic spin-cluster expansion including
  higher-order interactions\cite{Drautz04,Singer11} for a fit to non-collinear
  states.}

The results of the two approaches (infinitesimal rotations and
spin-wave spectra) on the values of exchange interactions agree well,
at least as long as the density-functional calculations are done
within the same electronic structure method.\cite{FrotaPessoa00}
\new{There is also good agreement in the spin-wave spectra of the two
  approaches if these are calculated in the first approach
  from a Fourier-transformation of the real space coupling parameters, as
  long as a sufficiently large number of atomic shells is taken in the
  Fourier series.\cite{Rusz06}}

The present paper contains two major parts. \new{In the first part,
  Sections \ref{sec:Jij} and \ref{sec:JijCalc}, we start by presenting
  the formalism for the calculation of pair exchange constants which
  we have implemented in the} full-potential augmented linearized
plane wave (FLAPW) method.\cite{Wimmer81,FLEUR} The approach of spin
spirals and inverse Fourier transformations\cite{Halilov98} is used to
arrive at formulae for the pair exchange constants in the case of one
or more magnetic atoms per unit cell. \new{Symmetry relations are
  derived that reduce the selection of spin-spirals to the irreducible
  part of the Brillouin zone. Then, we focus on} accuracy tests
concerning the use of the force theorem at finite rotations; we also
address the problem of subtracting the contribution of the
magnetization in the interstitial. As a test, we calculate the Curie
temperature of certain compounds by a Monte Carlo method. In the
second part, Section \ref{Sec:longitudinal}, we discuss a way to
parametrize the energetic contribution of longitudinal changes in the
atomic spin moment so as to include them in a Monte Carlo method in
the case that we are faced with a compound containing strong-moment
and induced-moment sublattices. We obtain a scheme for the study of
temperature-dependent magnetic properties and derive equations that
allow a simplification of the computational method and a reduction of
the computational cost under certain frequently-met physical
conditions; \new{importantly, we show the hitherto unnoticed result
  that the fluctuating} weak-moment degrees of freedom can be
\new{analytically} eliminated in favour of renormalized,
temperature-independent strong-moment interactions, \new{and that the
  thermodynamics of the full system (strong plus weak moments with
  bare interactions) can be derived from the thermodynamics of the
  strong-moment system only with renormalized interactions (a general
  derivation is given in the Appendix). A numerical demonstration of
  this rigorous result is presented in the case of the half-Heusler
  alloy NiMnSb. Finally, in Sec.~\ref{sec:remarks} we place our work in a
  wider context comparing with the treatment of the weak moments or
  the concept of renormalized exchange interactions presented by other
  authors and we state the basic physical assumptions underlying our
  result. We then conclude with a summary.}

\section{\textit{Ab-initio} Calculation of Heisenberg
Exchange Parameters \label{sec:Jij}}

Adopting the adiabatic approximation, the magnetic interactions are
modelled by a classical Heisenberg Hamiltonian. The part of the total
energy due to these interactions is then obtained from the expression
\begin{equation}\label{eq:heisenberg1}
E=-\frac{1}{2}\!\!\!\!\!\sum_{\stackrel{\stackrel{
n,m}{\alpha\beta}}
{(\vc{R}_{m\alpha}\neq\vc{R}_{n\beta})}}
\!\!\!\!\!
J(\vc{R}_{m\alpha},\vc{R}_{n\beta})\,
\vc{M}_{m\alpha}\cdot\vc{M}_{n\beta}
,
\end{equation}
where
$\vc{R}_{m\alpha(n\beta)}\equiv\vc{R}_{m(n)}+\vc{\tau}_{\alpha(\beta)}$. Here,
$\vc{R}_{m(n)}$ are the lattice vectors and
$\vc{\tau}_{\alpha(\beta)}$ are the basis vectors specifying the
positions of the atoms within the unit cell.
$\vc{M}_{m\alpha(n\beta)}$ are the atomic magnetic moments at the
sites $\vc{R}_{m\alpha(n\beta)}$, while
$J(\vc{R}_{m\alpha},\vc{R}_{n\beta})$ is the exchange coupling
constant for the pair of atoms situated at these sites, and is the
quantity to be calculated. The summations over the indices $n,m$ are
carried out over all lattice vectors, and the ones using indices
$\alpha,\beta$, over all the atoms in the unit cell. The factor 1/2
takes care of the double counting and the on-site term
($\vc{R}_{m\alpha}=\vc{R}_{n\beta}$) is left out.

The constants $J(\vc{R}_{m\alpha},\vc{R}_{n\beta})$ contain
the information about the inter-site interaction due to the exchange
coupling. The knowledge of these exchange interactions is essential
for the description of thermal excitations in magnetic solids and
their deriving from \textit{ab-initio} calculations is the core
problem in the attempt to describe the system with the Heisenberg
Hamiltonian. The correspondence between the \textit{ab-initio} theory
and the Heisenberg model is established by using the ansatz
\begin{equation}
J(\vc{R}_{m\alpha},\vc{R}_{n\beta}) =
\frac{\delta^2E}{\delta \vc{M}_{m\alpha} \delta \vc{
M}_{n\beta}} = \frac{\delta^2E_{\mathrm{DFT}}}{\delta \vc{
M}_{m\alpha} \delta \vc{M}_{n\beta}},
\label{eq:Jdef}
\end{equation}
which follows from Eq.~(\ref{eq:heisenberg1}), as a defining relation
of $J(\vc{R}_{m\alpha},\vc{R}_{n\beta})$ within density-functional
theory.  Here, $\delta \vc{M}_{n\alpha}$ and $\delta \vc{M}_{n\beta}$
are to be understood as small differences with respect to the
direction only, not the magnitude.  I.e., an appropriate energy
functional (usually within the local density or generalized gradient
approximation (LDA or GGA)) $E_{\mathrm{DFT}}[\rho,\vc{m}]$ of charge density
$\rho(\vc{r})$ and magnetization density $\vc{m}(\vc{r})$ is used in
the place of $E$ in Eq.~(\ref{eq:Jdef}). When evaluating
(\ref{eq:Jdef}), it is assumed that the exchange-correlation field is
collinear within each atom, so that the derivative with respect to
atom-cell integrated moments $\vc{
  M}_{m\alpha}=\int_{\mathrm{cell}\,m\alpha}\vc{m}(\vc{r})\,d^3r$ is
meaningful. The intra-atomic collinearity is an approximation,
justified by the energetic dominance of the moment formation (usually
of the order of eV) compared to the formation of ferromagnetic order
(usually of the order of less than 0.1 eV). From these comments it is
also evident that we do not require that the local moments are
quantized either in the form $M= \sqrt{S(S+1)}\ \muB$ or $M_z=S\ \muB$
with $S$ integer or half-integer, as, e.g., would be the case in
ferromagnetic semiconductors.\cite{Nolting79} Rather,
Eq.~(\ref{eq:heisenberg1}) represents the lowest term in an expansion
of the total energy in terms of the magnetization direction,
neglecting longitudinal enhancement or suppression of the moments, and
Eq.~(\ref{eq:heisenberg1}) represents a classical Heisenberg model,
valid after local quantum spin fluctuations have been averaged out
(see, e.g., the discussion in [\onlinecite{Halilov98}]). There are
known cases when these approximations are insufficient, and we discuss
such a case in Section \ref{Sec:longitudinal}.

Next, the collective transverse magnetic excitations are approximated
by static spin spirals, the energy of which is calculated within the
non-collinear FLAPW method.\cite{FLEUR,Kurz04} \new{For the Fourier and
back-Fourier transformations that are needed we follow the formalism
of Halilov {\it et al.}.\cite{Halilov98}} In the case of a spin spiral
with wave vector $\vc{q}$, the azimuthal angle of the magnetic moment
of an atom at the position $\vc{R}_{n\alpha}$ is given by
$\varphi_{n\alpha} = \vc{q} \cdot \vc{R}_{n\alpha}$. The magnetic
moment of an atom at the position $\vc{R}_{n\alpha}$ is
\begin{equation}
\label{eq:magnm}
  \vc{M}_{n\alpha} = M_{\alpha} \left( 
\begin{array}{l}
\sin \theta_{\alpha}\,\cos(\vc{q} \cdot
  \vc{R}_{n\alpha}+\phi_{\alpha}) \\
\sin \theta_{\alpha}\, \sin(\vc{q} \cdot
  \vc{R}_{n\alpha}+\phi_{\alpha}) \\
 \cos \theta_{\alpha} 
\end{array}
\right),
\end{equation}
where $\theta_{\alpha}$ is the so-called \textit{cone angle}, a
relative angle between the final and initial direction of the local
magnetic moment (here chosen along the $z$ axis; this choice does not
limit the generality in absence of spin-orbit coupling), and
$\phi_{\alpha}$ an eventual phase factor, also called \textit{phase
  angle}.  Taking advantage of the translational invariance we define
$\vc{R}\equiv\vc{ R}_{n} -\vc{R}_{m}$ and
$\boldsymbol{\tau}_{\alpha\beta}\equiv\boldsymbol{\tau}_{\alpha}-\boldsymbol{\tau}_{\beta}$,
whence Eq.~(\ref{eq:heisenberg1}) becomes
\begin{widetext}
\begin{eqnarray}\label{eq:heisenberg2}
E(\vc{q};\Theta;\Phi)&=&-\frac{1}{2}
\sum_{\stackrel{\alpha\beta}{\vc{R}}}'
M_{\alpha}M_{\beta}J(\boldsymbol{\tau}_{\alpha},\boldsymbol{\tau}_{\beta}
-\vc{R})
\nonumber\\
&\times&
\left\{\sin\theta_{\alpha}\sin\theta_{\beta}\cos\left[\vc{q}\cdot
(\boldsymbol{\tau}_{\alpha\beta}-\vc{R})+
\phi_{\alpha}-\phi_{\beta}\right] 
+\cos\theta_{\alpha}\cos\theta_{\beta}\right\}.
\end{eqnarray}
Here, the energy $E$ is a function of the spin-spiral vector $\vc{
  q}$, as well as of the cone and phase angles of the magnetic moments
on all the atoms of the unit cell. The dependence on these angles is
collectively expressed by $\Theta$ for the set of all cone angles
$\{\theta_\alpha\}$ and by $\Phi$ for the set of all phase angles
$\{\phi_\alpha\}$ in the argument of $E$.  To account for the
condition
$\boldsymbol{\tau}_{\alpha}\neq\boldsymbol{\tau}_{\beta}-\vc{R}$ under
which the sum in Eq.~(\ref{eq:heisenberg2}) is conducted (and is
indicated by a prime), from now on we set
$J(\boldsymbol{\tau}_{\alpha},\boldsymbol{\tau}_{\alpha})\equiv 0$,
for all the atoms $\alpha$ in the unit cell.

With the aim to obtain the exchange interaction constants
$J(\boldsymbol{\tau}_{\alpha},
\boldsymbol{\tau}_{\beta}-\vc{R})$ at the minimum of
computational expense, we define in the following a set of expressions
which are evaluated computationally. We first define the Fourier
transform
\begin{equation}\label{eq:Fourier}
J_{\alpha\beta}(\vc{q})= \sum_{\vc{R} }
J(\boldsymbol{\tau}_{\alpha},\boldsymbol{\tau}_{\beta}-
\vc{R}) \, e^{i\vc{q} 
\cdot(\boldsymbol{\tau}_{\alpha\beta}-\vc{R})}.
\end{equation}
It is straightforward to show that with the use of this Fourier
transform, Eq.~(\ref{eq:heisenberg2}) becomes
\begin{eqnarray}\label{eq:heisenberg3}
E(\vc{q};\Theta;\Phi) &=&
-\frac{1}{2}\sum_{\alpha\beta} M_{\alpha} M_{\beta} 
\left\{\sin\theta_{\alpha}\sin\theta_{\beta}{\mathcal Re}
\left[J_{\alpha\beta}(\vc{q})
e^{i(\phi_{\alpha}-\phi_{\beta})}\right] 
+\cos\theta_{\alpha}\cos\theta_{\beta}J_{\alpha\beta}(\vc{0})\right\}.
\end{eqnarray}
\end{widetext}

\subsection{Symmetry Relations}

Starting from the condition that $J(\vc{R}_{m\alpha},\vc{
R}_{n\beta})$ are real and symmetric and from the definition of the Fourier
transform $J_{\alpha\beta}(\vc{q})$ (Eq.~\ref{eq:Fourier}),
several useful symmetry relations of $J_{\alpha\beta}(\vc{q})$
can be derived (valid for each $\vc{q}$ vector):
\begin{enumerate}
\item $J_{\alpha\beta}(\vc{q})=J_{\beta\alpha}(-\vc{q})$
\item ${\mathcal Re}\left[J_{\alpha\beta}(\vc{q})\right]=
{\mathcal Re}\left[J_{\alpha\beta}(-\vc{q})\right]$
\item ${\mathcal Im}\left[J_{\alpha\beta}(\vc{q})\right]=
-{\mathcal Im}\left[J_{\alpha\beta}(-\vc{q})\right]$
\begin{itemize}
\item[(3{\textit a})] ${\mathcal Im}\left[J_{\alpha\beta}(\vc{0})\right]=0$
\item[(3{\textit b})] ${\mathcal Im}\left[J_{\alpha\alpha}(\vc{q})\right]=0$
\end{itemize}
\item  $J_{\alpha\beta}(\hat{C}\vc{q})=J_{\alpha'\beta'}(\vc{q})$,
where $\hat{C}$ is a crystal point group symmetry element and
$(\alpha',\beta')$ are the equivalent sites in the unit cell to
$(\alpha,\beta)$ via the action of the symmetry element
$\hat{C}^{-1}$, i.e.,
$\hat{C}^{-1}\boldsymbol{\tau}_{\alpha}=\boldsymbol{\tau}_{\alpha'}
+ \vc{R}$ for some lattice vector $\vc{R}$ (and analogously for $\beta$).
\end{enumerate}
\new{Symmetry Relations 1-3 have been given previously in
  Ref.~\onlinecite{Halilov98}.} 
Symmetry Relation 4 has the important consequence  that the
$\vc{q}$ vectors can be sampled from the irreducible wedge of the
Brillouin zone while $J_{\alpha\beta}(\vc{q})$ in the rest of the
Brillouin zone can be obtained by the symmetry transformations. In
case that the crystal possesses the inversion symmetry $\hat{I}$ and
if $\boldsymbol{\tau}_\alpha-\hat{I}\boldsymbol{\tau}_\alpha\equiv
2\boldsymbol{\tau}_\alpha$ and
$\boldsymbol{\tau}_\beta-\hat{I}\boldsymbol{\tau}_\beta\equiv
2\boldsymbol{\tau}_\beta$ are both lattice vectors, then from Symmetry
Relations 3 and 4 it follows that all $J_{\alpha\beta}(\vc{q})$ are
real.  Moreover, due to Symmetry Relation 1, even if the system does
not possess the inversion symmetry it is not necessary to make two
separate calculations for $\vc{q}$ and $-\vc{q}$.

\subsection{Calculational Scheme}
To develop a scheme for the calculation of the Fourier transforms
$J_{\alpha\beta}(\vc{q})$, we distinguish two different cases in the
calculational setup.

\noindent
$\bullet$\textbf{Case 1:} All the atoms in the unit cell are ordered
in the collinear ground state, except for atom $\mu$. Its magnetic
moment is tilted by the cone angle $\theta_\mu$ and the spin spiral running
through the system will affect only the magnetic moments situated on the
atoms of the same kind as $\mu$. In short: 
\begin{eqnarray*}
\theta_\mu &\neq& 0 \\
\theta_\lambda&=&0,\, \forall\lambda\neq\mu
\end{eqnarray*}
\setlength{\unitlength}{1cm} \mbox{\begin{pspicture}(2.6,0.8)
    \psframe(0,-0.2)(2.4,0.6)\psline{->}(0.3,0)(0.3,0.4)\psline{->}(0.6,0)(0.6,0.4)\psline{->}(0.9,0)(0.9,0.4)
    \psset{linecolor=red}{\psline{->}(1.2,0)(1.4,0.38)}\psset{linecolor=black}\psline{->}(1.5,0)(1.5,0.4)
    \psline{->}(1.8,0)(1.8,0.4)\psline{->}(2.1,0)(2.1,0.4)
\end{pspicture}}\\ \ \\

With the use of the symmetry relations for
the coefficients $J_{\alpha\beta}(\vc{q})$, from the total
energy expression (\ref{eq:heisenberg3}) one obtains
\begin{equation}\label{eq:Jmm}
J_{\mu\mu}(\vc{q})-J_{\mu\mu}(\vc{0})=-2\frac{E^\mu(\vc{q})
-E^\mu(\vc{0})}{M^{2}_{\mu}\sin^2\theta_{\mu}}.
\end{equation}

\noindent
$\bullet$\textbf{Case 2:} 

This case will appear only if there are two or more magnetic atoms in
the unit cell.  Keeping the rest of the magnetic moments parallel, the
magnetic moments on atoms $\mu$ and $\nu$ are tilted by cone angles
$\theta_\mu$ and $\theta_\nu$, respectively, so the spin spiral
running through the system changes the orientation of magnetic moments
on both of these atoms:
\begin{eqnarray*}
\theta_\mu,\theta_\nu&\neq& 0\\ 
\theta_\lambda&=&0,\, \forall\lambda\neq\mu,\nu
\end{eqnarray*}
\setlength{\unitlength}{1cm}
\mbox{\begin{pspicture}(2.6,0.8)
\psframe(0,-0.2)(2.4,0.6)\psline{->}(0.3,0)(0.3,0.4)\psset{linecolor=red}\psline{->}(0.6,0)(0.4,0.4)\psset{linecolor=black}
\psline{->}(0.9,0)(0.9,0.4)
\psset{linecolor=red}{\psline{->}(1.2,0)(1.4,0.38)}\psset{linecolor=black}\psline{->}(1.5,0)(1.5,0.4)
\psline{->}(1.8,0)(1.8,0.4)\psline{->}(2.1,0)(2.1,0.4)
\end{pspicture}}\\ \ \\

As we have seen, if the system does not
possess inversion symmetry, the coefficients $J_{\mu\nu}(\vc{q})$
are complex for $\mu\neq\nu$. Their real and imaginary part can be
obtained as
\begin{widetext}
\begin{eqnarray}
{\mathcal Re}\left[J_{\mu\nu}(\vc{q})\right] &=&
\frac{1}{M_{\mu}M_{\nu}\sin\theta_{\mu}\sin\theta_{\nu}} 
\left\{
E^{\mu\nu}(\vc{0},\frac{\pi}{2})
-E^{\mu\nu}(\vc{q},0) 
+E^{\mu}(\vc{q}) - E^{\mu}(\vc{0})+ 
E^{\nu}(\vc{q})-E^{\nu}(\vc{0}) \right\} \label{eq:realJ} 
\\ 
{\mathcal Im}\left[J_{\mu\nu}(\vc{q})\right] &=&
\frac{E^{\mu\nu}(\vc{q},\frac{\pi}{2})
-E^{\mu\nu}(\vc{q},0)}
{M_{\mu}M_{\nu}\sin\theta_{\mu}\sin\theta_{\nu}}-{\mathcal
  Re}\left[J_{\mu\nu}(\vc{q})\right]. \label{eq:imaginaryJ}
\end{eqnarray}
where $E^{\mu\nu}(\vc{q},0)$ and $E^{\mu\nu}(\vc{
q},\frac{\pi}{2})$ denote the total energies in the presence of a
spin spiral defined with the wave vector $\vc{q}$ and the
difference of the phase factors $\phi_{\mu}-\phi_{\nu}=0$ and
$\frac{\pi}{2}$ respectively.

\subsection{Brillouin Zone Integration}

We have established that the Fourier transforms
$J_{\mu\nu}(\vc{q})$ can be obtained from the differences in
total energy between the states having specified magnetic
configurations. Armed with Eqs.~(\ref{eq:Jmm},~\ref{eq:realJ})
and~(\ref{eq:imaginaryJ}) we are now ready to calculate the Heisenberg
exchange coupling constants,
$J(\boldsymbol{\tau}_{\mu},\boldsymbol{\tau}_{\nu}- \vc{
R})$. First, however, one has to take into account that from the
Eq.~(\ref{eq:Jmm}) it is only possible to calculate the difference
$J_{\mu\mu}(\vc{q})-J_{\mu\mu}(\vc{0})$, but not the
coefficient $J_{\mu\mu}(\vc{q})$ alone. This problem can be
easily circumvented by introducing the coefficients
$\tilde{J}_{\mu\nu}(\vc{q})$, defined as
\begin{equation}\label{eq:Jtilde}
\tilde{J}_{\mu\nu}(\vc{q})\equiv J_{\mu\nu}(\vc{q})-\delta_{\mu\nu}J_{\mu\nu}(\vc{0}).
\end{equation}
Also, for simplicity, the non-zero cone angles can in all calculations
be taken to have the same value $\theta$.
Eqs.~(\ref{eq:Jmm},~\ref{eq:realJ}) and~(\ref{eq:imaginaryJ}) can now
be re-written as
\begin{eqnarray}
\tilde{J}_{\mu\mu}(\vc{q})&=&-2\,\frac{E^\mu(\vc{q})
-E^\mu(\vc{0})}{M^{2}_{\mu}\sin^2\theta}\label{eq:Jtilde_MM}\nonumber\\
{\mathcal Re}\left[\tilde{J}_{\mu\nu}(\vc{q})\right]
&=&\frac{E^{\mu\nu}(\vc{0},\frac{\pi}{2})
-E^{\mu\nu}(\vc{q},0)}
{M_{\mu}M_{\nu}\sin^2\theta} -\frac{1}{2}\frac{M_{\mu}}{M_{\nu}}\tilde{J}_{\mu\mu}-
\frac{1}{2}\frac{M_{\nu}}{M_{\mu}}\tilde{J}_{\nu\nu}\label{eq:ReJtilde}\nonumber\\
{\mathcal Im}\left[\tilde{J}_{\mu\nu}(\vc{q})\right]&=&\frac{E^{\mu\nu}(\vc{q},\frac{\pi}{2})
-E^{\mu\nu}(\vc{q},0)}
{M_{\mu}M_{\nu}\sin^2\theta}-{\mathcal Re}\left[\tilde{J}_{\mu\nu}(\vc{q})\right].\label{eq:ImJtilde}
\end{eqnarray}
\end{widetext}
The final step will be a simple back-Fourier transform. Using
Eqs.~(\ref{eq:Fourier}) and (\ref{eq:Jtilde}) it is easy to see that
\begin{equation}
\label{eq:integration}
J(\boldsymbol{\tau}_{\mu},\boldsymbol{\tau}_{\nu}-\vc{R})=
\frac{1}{V_{BZ}}\int_{V_{BZ}}\tilde{J}_{\mu\nu}(\vc{q}) \,
e^{-i\vc{q}\cdot(\boldsymbol{\tau}_{\mu\nu}-\vc{R})}\,d^3q.
\end{equation}
Finally, we note that from the definition~(\ref{eq:Jtilde}) it is clear
that $\tilde{J}_{\mu\nu}(\vc{q})$ satisfies the same symmetry
relations as the coefficients $J_{\mu\nu}(\vc{q})$.

The described calculations can be time consuming, since they involve
the determination of small energy differences (typically of the order
of a few mRyd). Due to the oscillatory phase in
Eq.~(\ref{eq:integration}), the appropriate size of the $\vc{q}$ point
set increases with the distance between the two atoms for which the
interaction constant is being calculated, the grid fineness being
basically determined by the inverse of the quantity
$|\boldsymbol{\tau}_{\mu\nu}-\vc{R}|$ that enters the exponential in
Eq.~(\ref{eq:integration}). In the spirit of the Nyquist-Shannon
sampling theorem,\cite{Shannon49} and for
$|\vc{R}|\gg|\boldsymbol{\tau}_{\mu\nu}|$, the $\vc{q}$-grid fine
spacing in the direction of $\vc{R}$ should be at most half the value
$2\pi/|\vc{R}|$ . Additionally, sufficient accuracy requires larger
plane-wave basis sets and a finer $\vc{k}$-point grid compared to a
simple ground-state calculation. A rule of thumb for increased accuracy is
that, given a $\vc{q}$-grid the $\vc{k}$-grid should be twice as fine
per spatial dimension of the lattice in order to avoid spurious
oscillatory behaviour of period of the grid-spacing $\delta\vc{k}$ in
$\tilde{J}_{\mu\nu}(\vc{q})$. A self-consistent calculation of all
energies needed here is computationally very demanding.  Fortunately,
in many cases the spin spiral can be considered a small enough
perturbation that the force theorem \cite{Andersen,Oswald85} can be used
to calculate the energy differences. We discuss this in the following
subsection.

\subsection{Test of the Applicability of the Force Theorem}

The magnon energy is calculated as the difference between the total
energy of the system with a spin spiral (this is an excited state),
and the ground state, which is ferromagnetic in the systems under
study here. A self-consistent calculation of the spin spiral requires
use of a constraint, in the form of an external spiraling magnetic
field, which forces the magnetization to take the form given in
Eq.~(\ref{eq:magnm}). On the other hand, an application of the force
theorem requires a position dependent rotation of the
exchange-correlation field $\vc{B}^{xc}$ so that its direction
acquires the form (\ref{eq:magnm}), i.e.,
\begin{equation}
\label{eq:bnoco}
  \vc{B}_{n\alpha}^{xc} = B_{\alpha}^{xc} \, \left( 
\begin{array}{l}
\sin \theta_{\alpha} \cos(\vc{q} \cdot
  \vc{R}_{n\alpha}+\phi_{\alpha}) \\
 \sin \theta_{\alpha} \sin(\vc{q} \cdot
  \vc{R}_{n\alpha}+\phi_{\alpha})\\ 
\cos \theta_{\alpha} 
\end{array}
\right),
\end{equation}
where
$B_{\alpha}^{xc}=(V_{\uparrow\,\alpha}^{xc}-V_{\downarrow\,\alpha}^{xc})$
is the self-consistent exchange-correlation field of the collinear
calculation at atom type $\alpha$ ($V_{\sigma\alpha}^{xc}$ is the
exchange-correlation potential dependent on spin
$\sigma=\uparrow,\downarrow$). In the FLAPW method, this rotation is
applied in the muffin-tin spheres, i.e., non-overlapping spheres
around the atomic nuclei where the potential is expanded in radial
functions and spherical harmonics, \new{as well as in the interstitial}
 space between these spheres. Using the field of
Eq.~(\ref{eq:bnoco}) in the Kohn-Sham equations yields a
(non-self-consistent) sum of eigenvalues of the occupied levels;
according to the force theorem, the magnon energy is approximated by
the difference of this sum to the sum of eigenvalues of occupied
levels in the self-consistent ferromagnetic ground state.

\begin{figure}[h!]
\includegraphics*[width=7cm]{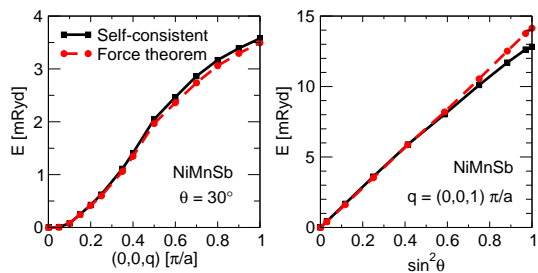}
\caption{(color online) Comparison of the force theorem with
  self-consistent calculations for NiMnSb. Left: a dispersion curve of
  a spin spiral along the [001] direction for a cone angle
  $\theta=30^{\rm o}$. Right: spin spiral energy vs. $\sin^2\theta$
  for a short-wavelength spiral along the [001]
  direction.\label{fig:force}}
\end{figure}

The approximation is expected to be better for smaller perturbations,
i.e., for smaller cone angles and/or smaller magnon wavevectors
$|\vc{q}|$. As a test, in the left panel of Figure~\ref{fig:force} the
dispersion curve is shown for a spin spiral in NiMnSb, defined by a
cone angle $\theta=30^{\rm o}$, and a spin-wave vector along [001]
direction. The force-theorem and self-consistent calculations agree
rather well. The right panel of Fig.~\ref{fig:force} shows the
dependence of the magnon energy on the squared sinus of the cone angle
($\sin^2\theta$) for a fixed spin-spiral vector $\vc{q}=(0, 0,
1)\pi/\alat$. We see a a maximal deviation of the order of 7-8\% for
the unfavourable case of $\theta=90^{\rm o}$ and $q=\pi/\alat$, while
the deviation starts becoming visible at cone angles larger than
$\theta_{\rm max}\sim 50^{\rm o}$.

A general conclusion is that if one wants to use the force theorem to
obtain the spin-spiral dispersion within the whole first Brillouin
zone of the crystal, a cone angle $\theta=30^{\rm o}$ seems to be a
reasonable choice, since the energy differences are not too large for
the magnon to stop being a small perturbation, but are also not too
small so that one would have to employ a very large basis, or
\vc{k}-point set.  This cone angle was used in the calculation of
the Heisenberg interaction constants $J$ presented in
Section~\ref{sec:JijCalc}.

\begin{figure*}
\includegraphics*[width=17cm]{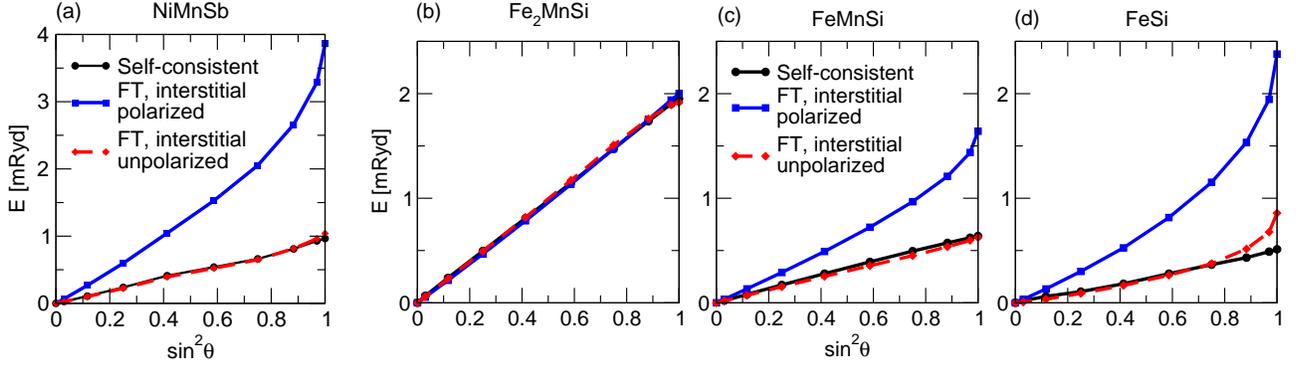}
\caption{(color online) (a): Spiral energy $E(\vc{q},\theta)$ as function of
  $\sin^2\theta$ in NiMnSb for the cases of  (i) a self-consistent
  calculation (including the self-consistent, spiraling polarization
  of the interstitial) (ii) a force-theorem
  calculation including the polarization of the
  interstitial, and (iii) a force-theorem calculation excluding the
  polarization of the interstitial. Evidently, case (ii) stongly overestimates the
  spiral energy, since the interstitial polarization
  acts as an effective magnetic field. Case (iii) on the other hand is
  a good approximation to the exact result (i). The spiral
  wavevector is $\vc{q}=(0,0,0.15)\,(\pi/\alat)$. (b,c,d): Same as in
  (a) but for Fe$_2$MnSi, FeMnSi, and FeSi, in full-Heusler,
  half-Heusler, and zinc-blende structures, respectively, to
  demonstrate the effect of increasing the volume of the
  interstitial. The structures correspond to an fcc lattice with four,
  three, and two atoms per unit cell, respectively. All three were
  calculated in the same lattice parameter ($\alat=5.663$~\AA) and at a
  spiral vector of $\vc{q}=(0,0, 0.25)\,(\pi/\alat)$.
  \label{fig:interstitial}}
\end{figure*}
We close this section with a comment on a spurious effect of the
exchange-correlation field in the interstitial region when the force
theorem is used. \new{(The interstitial region is covered by empty
  spheres or empty cells in some ab-initio methods, e.g. in the
  ASW\cite{Sandratskii02,Kubler06} or LMTO method,\cite{Halilov98} if
  the structures are relatively open, as are for instance the
  half-Heusler or the zinc-blende structure.)} In a force-theorem
application, the trial exchange-correlation potential in the
interstitial is a smooth periodic function. However, in a
self-consistent spin-spiral calculation, the resulting
exchange-correlation potential is in many cases a much less smooth,
although still periodic, function.\cite{Bylander98} Then the
interstitial magnetization in the force-theorem calculation can cause
a serious overestimation of the spin-spiral energy. This spurious
energy contribution can be circumvented by setting the magnetic part
of the exchange-correlation potential in the interstitial to
zero. Depending on the volume filling of the touching ``muffin-tin''
atomic spheres of the system, the spurious energy can be considerable,
becoming larger for open systems. In Fig.~\ref{fig:interstitial}a we
show a calculated example for NiMnSb. The self-consistent spiral
energy and the force-theorem energy (here normalized to
$\sin^2\theta$) are practically identical, if we set $\vc{B}^{xc}=0$
in the interstitial in the force theorem calculation (in the
self-consistent calculation, $\vc{B}^{xc}$ is non-zero also in the
interstitial); if this correction is not used, then the spiral energy
is strongly overestimated. Figs.~\ref{fig:interstitial}b-d show the
same for Fe$_2$MnSi in the full-Heusler structure, FeMnSi in the
half-Heusler structure, and FeSi in the zincblende structure. All
three structures are based on an fcc lattice and were calculated here
with the same lattice parameter $\alat=5.663$~\AA, but differ in the
number of atoms in the unit cell and therefore in the volume of the
interstitial region. Fe$_2$MnSi has the smallest interstitial region,
and evidently the interstitial magnetization has almost no effect;
both force theorem calculations practically coincide with the
self-consistent result. For FeMnSi, however, the interstitial volume
is larger and the spiral energy is strongly overestimated in the force
theorem calculation, if the interstital magnetization is not set to
zero. The effect is even stronger for FeSi, where the interstitial
volume is largest.

\section{Exchange Interaction Parameters and Curie Temperature\label{sec:JijCalc}}

\begin{figure}
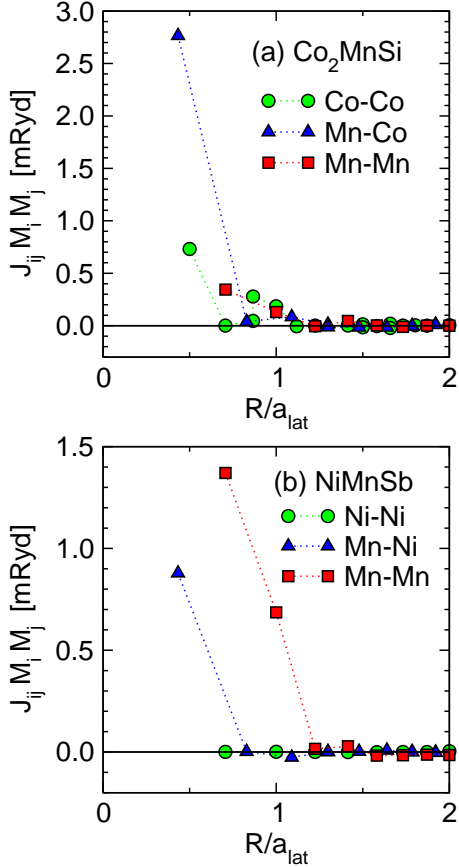

\begin{center}
\includegraphics*[width=6cm]{jCo2MnSi.eps}
\includegraphics*[width=6cm]{jNiMnSb.eps}
\end{center}
\caption{(color online) Heisenberg exchange interaction parameters of
Co$_2$MnSi (a) and NiMnSb (b) as a function of the distance R between
the atoms (in units of the lattice constant~\alat). The lines are guides
to the eye.\label{fig:exchange}}
\end{figure}

Following the prescription of Sec.~\ref{sec:Jij} we calculated
Heisenberg exchange interaction parameters ($J_{ij}$; $i,j$ are
abbreviations of ${m\alpha},{n\beta}$) of Co$_2$MnSi and NiMnSb, shown
in Fig.~\ref{fig:exchange} and used a Monte Carlo method to obtain an
estimate of the Curie temperatures for these compounds. The
calculations for both compounds were performed within the generalized
gradient approximation\cite{PBE} on a 4096 $\vc{k}$-point mesh with 2744
$\vc{q}$-points in the full Brillouin zone.  The planewave cutoff was
$k_{\mathrm{max}}=3.8\ \mathrm{au}^{-1}$. The convergence was checked
with respect to the above parameters.

\begin{figure}[h]
\begin{center}
\includegraphics*[width=6cm]{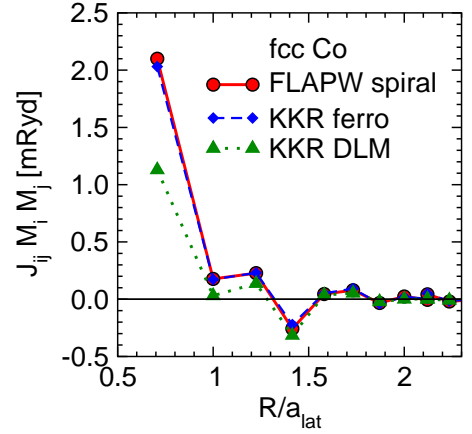}
\end{center}
\caption{Heisenberg exchange interaction parameters of fcc~Co
as a function of the distance R between the atoms (in units of the
lattice constant~a). The lines are guides to the eye. Red circles:
results calculated with the FLAPW method. Blue diamonds: results
calculated with the Korringa-Kohn-Rostoker (KKR) Green function method
within the approximation of infinitesimal rotations with the
ferromagnetic state as reference. Green triangles: results calculated
again with KKR but with the disordered local moment (DLM) state as reference. 
\label{fig:Co}}
\end{figure}

\begin{figure}[h]
\begin{center}
\includegraphics*[width=7cm]{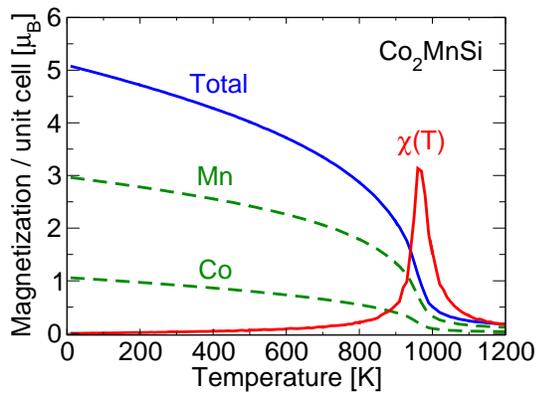}
\end{center}
\caption{(color online) Magnetic moment $M(T)$ per unit cell (blue
  full line), partial moment of the Mn and Co sublattices (green
  dashed lines), and susceptibility $\chi(T)$ (red full line) as
  functions of the temperature $T$, for Co$_2$MnSi. The peak of the
  susceptibility at 980~K indicates the Curie temperature
  (experimentally found to be 985 K). Note that the susceptibilities
  are shown in an arbritary scale.
\label{fig:MCclassic}}
\end{figure}

In Co$_2$MnSi, both Co and Mn atoms have strong and stable magnetic
moments, whose interaction is described with $J$ parameters depicted
in Fig.~\ref{fig:exchange}a.  In NiMnSb, though the magnetic moment of
Ni is small and actually induced by the Mn surrounding atoms, both Ni
and Mn were treated as magnetic atoms and the parameters of Mn-Mn,
Ni-Ni, and Mn-Ni interaction were calculated
(Fig.~\ref{fig:exchange}b). As will be discussed in
Sec.~\ref{Sec:longitudinal}, this treatment gives an insight into the
thermal behaviour of the Ni sublattice and with the use of an extended
Heisenberg model more useful information can be obtained.

Co$_2$MnSi and NiMnSb are half-metallic ferromagnets, i.e., the
density of states (DOS) in one spin direction (here majority spin) is
metallic, while in the other spin direction the DOS shows a band gap
around the Fermi level. 

\new{For a non-half-metallic ferromagnet, the interaction constants
  $J_{ij}$ as a function of
  distance follow a decaying oscillating behavior that in the simplest case is of the
  Ruderman-Kittel-Kasuya-Yoshida (RKKY) type, decaying with
  $|\vc{R}_i-\vc{R}_j|^{-3}$, but in general can have many
  superimposed periods and a different decay power law depending on the
  details of the Fermi surface.\cite{Pajda01} On the other hand, for
  half-metallic ferromagnets the gap at one spin direction leads to an
  imaginary wave vector and an exponential decay with
  distance.\cite{Pajda01} } In both cases shown in
Fig.~\ref{fig:exchange}, we notice this very fast decay of the
interaction parameters with the distance between the atoms. For a
comparison, in Fig.~\ref{fig:Co}, the exchange interaction parameters
of fcc~Co (which is not half-metallic) are shown.  The decay here is
much slower.


\new{In a short digression, we compare the results of the spin-spiral
  approach within the FLAPW method to the approach of infinitesimal
  rotations\cite{Liechtenstein87} calculated with the full-potential
  Korringa-Kohn-Rostoker Green function method (KKR).\cite{KKR} In KKR
  we use both the ferromagnetic state and the disordered local moment
  (DLM) state as reference points (see also the discussion in
  Sec.~\ref{sec:remarksII}).  We see in Fig.~\ref{fig:Co} that the
  agreement is good for fcc Co, with a deviation of 3\%
  nearest-neighbor coupling, if the ferromagnetic state is used as a
  reference, while the deviation is large if the DLM state is used as
  reference (also the atomic moment decreases from 1.65~\muB\ in
  the ground state to 1.08~\muB\ in the DLM state). For Co$_2$MnSi
  (not shown here) we find that the difference between KKR and FLAPW
  is larger, with the Mn-Co nearest-neighbor coupling being
  approximately 10\% weaker in the KKR calculation, while if the DLM
  state is used as reference, the Co moment vanishes altogether.}

Once the exchange interactions are known, the effective Heisenberg
model can be solved for the Curie temperature. While the mean field
approximation is computationally the simplest method to this task, it
is known (and has been shown in practice) that the resulting \Tc\ is
overestimated. The random phase approximation (also known as Tyablikov
approximation), on the other hand, is rather
accurate.\cite{RPA} The most accurate, but numerically more
expensive, way to calculate \Tc\ of a classical Heisenberg model is the
Monte Carlo method, in particular when taking advantage of the
cumulant expansion to account for the finite size of the simulation
supercells. In this work we applied the Monte Carlo method, locating
\Tc\ by the peak in the temperature-dependent, static
susceptibility. As the simulation supercells are rather large,
finite-size corrections to \Tc\ are small.

We also show Monte Carlo magnetization curves for Co$_2$MnSi in
Fig.~\ref{fig:MCclassic}.\cite{Mersene} The classical Heisenberg Hamiltonian,
Eq.~(\ref{eq:heisenberg1}) was used to model the systems, assuming
that the magnetic moments can change their orientation, but not their
length. NiMnSb is discussed in more detail in
Sec.~\ref{Sec:longitudinal}.  In the calculations the Metropolis
method was employed. A supercell of 2048 unit cells (4096 magnetic
atoms) was used for NiMnSb and one of 2197 unit cells (6591 magnetic
atoms) for Co$_2$MnSi; interactions to neighbors up to a distance of
four lattice constants were included. For each temperature the number
of Monte Carlo sampling events was 5000, after allowing an initial
relaxation time of 1000 steps and taking one sampling event every 10
sweeps of the lattice.

The magnetization curves (Fig.~\ref{fig:MCclassic}) do not go to a
sharp zero at \Tc, but rather have a tail, as a result of the finite
supercell. The peak of the susceptibility is, on the other hand,
rather sharp and its position can be used to determine the Curie
temperature. From the positions of these peaks, we estimated for
Co$_2$MnSi (Fig.~\ref{fig:MCclassic}, left) \Tc=980~K
(experimental\cite{Landolt} 985~K), and for NiMnSb (see
Sec.~\ref{Sec:longitudinal}) \Tc=870~K (experimental\cite{Landolt}
value being 730~K). Note that here, for NiMnSb, the Ni moment was also
taken into account within the Heisenberg model (but see also the
discussion in Sec.~\ref{Sec:longitudinal}, in particular
\ref{sec:NiMnSb}). For fcc Co we find a Curie temperature of 1200~K,
while the experimental result is 1403~K (the high-temperature stable
phase of Co is fcc).

\section{Consideration of Longitudinal Moment Fluctuations \label{Sec:longitudinal}}

So far we have discussed a formalism and examples of calculation of
exchange parameters for a Heisenberg model, assuming that the
magnitude of the local magnetic moments is rigid. However, as is long
known, certain systems are weakly magnetic and the moment formation
takes place at a relatively low energy scale, comparable to the magnon
energies. Moriya's quantum mechanical spin fluctuation
theory\cite{Moriya79} already factors in these effects. An
introduction of a classical Hamiltonian including longitudinal and
transverse degrees of freedom has been done by Uhl and
K\"ubler\cite{Uhl96} by parametrizing {\it ab-initio} total energy
results; thermodynamic quantities were then obtained within a
mean-field approach which coupled longitudinal and transverse degrees
of freedom.\cite{Uhl96,Kubler06} More elaborate approaches by a
classical fit of the energy to density-functional results, including
transversal and longitudinal degrees of freedom, together with
Monte-Carlo or classical spin-dynamics calculations, were presented
e.g. by \new{Rosengaard and Johansson,\cite{Rosengaard97}} Ruban {\it et al.},\cite{Ruban07} Ma and Dudarev,\cite{Ma12}
or Derlet.\cite{Derlet12}

Particularly interesting are compounds of a strongly magnetic and a
weakly magnetic subsystem, where the weak moments cannot be
treated as having rigid-magnitude, but on the other hand are large
enough that they cannot be neglected. Examples are FePt, FePd, or
NiMnSb, where the strongly magnetic atoms are Fe and Mn, while the
weakly magnetic atoms are Pt, Pd and Ni. Then the assumption of rigid
Heisenberg spins does no longer seem plausible. One way to circumvent
this is to use only the strong magnetic moments as independent
variables, but with their pair exchange parameters renormalized by the
exchange among the weakly magnetic atoms and by the enhanced
susceptibility. Such an approach was, for example, used by Mryasov
{\it et al.}\cite{Mryasov05,Mryasov05b} to model the temperature dependence of
the magnetic anisotropy in FePt. Sandratskii {\it et
  al.}\cite{Sandratskii07} also find improvement in the theoretical
results if the weak moments are not treated as independent, rigid
Heisenberg spins, but rather fully adjust to the magnetization of the
strong-moment sublattice. There is, however, the point of view of
treating the weak moments not as fully enslaved to their strong-moment
environment, but as independent degrees of freedom whose fluctuation
can affect the thermodynamics. This point of view has been pushed for
example in order to understand the antiferromagnetic-to-ferromagnetic
transition of FeRh.\cite{Gruner03,Sandratskii11} 

In this section we develop a method of treating such systems,
emphasizing the choise of the energetically relevant magnetic degrees
of freedom at temperatures up to the critical temperature, as well as
the practical implementation of these degrees of freedom to models
that can be solved via Monte Carlo simulations.  The additional
challenge, compared to the theory developed in the previous sections,
stems from two facts. First, a treatment of the weak moments as
independent degrees of freedom must include the energy scale of the
moments magnitude as well as their direction. Second, if the
interactions of the strong moments are calculated as described
earlier, by considering non-collinear magnetic configurations, then
the calculated energies include a spurious contribution because the
magnitude of the weak moments changes during the application of the
non-collinear constraint. This spurious contribution must be accounted
for. However, at the end we see that, in certain commonly occuring
situations, the degrees of freedom of the weak-moment atoms can be
eliminated in the statistical-mechanical calculations by using
renormalized exchange parameters, while the statistical averages of
the weak moments follow directly from the averages of the strong
moments. We show that this is an exact result if the weak degrees of
freedom give only quadratic contributions to the energy. Here we make
a step-by-step approach for the special case where the weak moments do
not interact with each other, while we present the more general case
when they interact in the Appendix.

\subsection{General approach \label{subsec:general}}

The magnetically strong atoms are the main carriers of spin moment
that also acts as an effective field $\vc{H}^{\rm eff}$ polarizing
the magnetically weak atoms. The polarization of the latter is
strengthened by an enhanced local susceptibility, which (neglecting
complications from the band-structure) is $\chi=\chi_{\rm
  Pauli}/(1-I\chi_{\rm Pauli})$ (with $\chi_{\rm Pauli}$ the Pauli
paramagnetic susceptibility and $I$ the exchange integral). Let us
denote by $\vc{M}$ the rigid-magnitude local moment of the
magnetically strong atoms and by $\vc{\mu}$ the supple local moment of
the magnetically weak atoms. A simple expression that parametrizes the
energy in terms of the local moment $\vc{\mu}$ and
the polarizing field $\vc{H}^{\rm eff}$ is
\begin{equation}
\Eloc[\vc{H}^{\rm eff},\vc{\mu}] = -\vc{H}^{\rm eff}\cdot\vc{\mu} +
a\mu^2 + b\mu^4 \label{eq:L1}
\end{equation}
It is implied that $\vc{H}^{\rm eff}=\kappa\sum_{n\in\rm
  neighb}\vc{M}_n + \vc{H}^{\rm ext}$, where $\vc{M}_n$ is the summed
moment of the neighboring atoms inducing a polarization, $\kappa$ is a
phenomenological parameter which encapsulates all microscopic
processes (in particular electron hoppings) that couple $\vc{\mu}$ to
$\vc{M}_n$, while $\vc{H}^{\rm ext}$ is an external magnetic field
that we henceforth set to zero.\cite{footnote} In the absence of a field
$\vc{H}^{\rm eff}$ Eq.~(\ref{eq:L1}) contains only even powers of
$\vc{\mu}$, because symmetry requires $\Eloc[0,\vc{\mu}]=\Eloc[0,-\vc{\mu}]$.

In Eq.~(\ref{eq:L1}) we always must have $b\geq 0$. The case of
spontaneous polarization is described by $a<0$ and $b>0$, while the
case of induced magnetic moments is described by $a>0$. In this case,
the on-site susceptibility is 
\begin{equation}
\chi = \partial \mu/\partial H^{\rm  eff}|_{H^{\rm eff}=0} = 1/(2a).
\label{eq:L1c}
\end{equation} 
Thus the enhancement of the susceptibility is contained in the
parameter $a$.

Eq.~(\ref{eq:L1}), if the fourth-order term is negligible, can be
written as
\begin{equation}
\Eloc[\vc{H}^{\rm eff},\vc{\mu}]= a\biggl(\vc{\mu}-\frac{1}{2a}\vc{H}^{\rm eff}\biggr)^2-\frac{1}{4a}(H^{\rm eff})^2,\ 
\text{if $b=0$.} 
\label{eq:L1b}
\end{equation}

To elaborate on these ideas we use NiMnSb as a concrete example. In
this case the Mn subsystem is magnetically strong and the Ni subsystem
magnetically weak. We have also found that in this case it is a good
approximation to set $b=0$ (as we show below), which we
adopt. Combining Eq.~(\ref{eq:L1}) at $b=0$ with the Heisenberg-model energy
expression for the Mn subsystem results in the extended model
Hamiltonian
\begin{eqnarray}
\mathcal{H}(\{\vc{M}_i;\vc{\mu}_l\}) &=& -\frac{1}{2}\sum_{ij \in \Mn} J^b_{ij}
\vc{M}_i\cdot\vc{M}_j  \nonumber\\
 &+& \sum_{l \in \Ni}
[a\mu_l^2-\kappa \vc{\mu}_l\cdot\sum_{n(l)}\vc{M}_n ]
\label{eq:L2}
\end{eqnarray}
where $\sum_{n(l)}\vc{M}_n$ is the sum of the moments of the Mn
nearest neighbors of the $l$-th Ni atom, $n(l)$, and $\vc{M}_i$ and
$\vc{\mu}_l$ refer to the Mn and Ni moment respectively. The
superscript ``$b$'' in the Mn-Mn exchange interaction is placed in
anticipation that it does not coincide with the quantity calculated
within the method presented in Sec.~\ref{sec:Jij}, but with a
``bare'' quantity, as will be discussed below. It has been also
assumed that the Ni-Ni interaction can be neglected, as there are no
Ni-Ni nearest neighbor pairs, while from the calculations it becomes
evident that more distant Ni-Ni interactions are negligible (the case
where the interactions among the weak-sublattice moments are
non-negligible is discussed in the Appendix). The above
equation can be rewritten in a somewhat more convenient form, if we
define (with an obvious notation)
\begin{eqnarray}
J^b_{\Mn-\Ni} &=& \left\{
\begin{array}{ll}
\kappa, & \text{nearest neighbours Mn-Ni} \\
0,  & \text{otherwise}
\end{array} \right. \label{eq:L2.5}\\
J_{\Ni-\Ni} &=& 0
\end{eqnarray}
Then Eq.~(\ref{eq:L2}) takes the form
\begin{eqnarray}
\mathcal{H} &=& -\frac{1}{2}\bigg[
\sum_{ij \in \Mn} J^b_{ij} \vc{M}_i\cdot\vc{M}_j +  
\sum_{i\in  \Mn,l\in \Ni} J^b_{il} \vc{M}_i\cdot\vc{\mu}_l  \nonumber\\
&&+\sum_{j\in \Mn,l\in \Ni} J^b_{lj} \vc{\mu}_l\cdot\vc{M}_j
 \bigg] + \sum_{l \in \Ni} a\mu_l^2\nonumber\\
\label{eq:L3}
\end{eqnarray}
This form is appealing, as it contains a Heisenberg-like expression
among all magnetic atoms plus a local correction term that accounts
for the longitudinal degree of freedom of the Ni moments. However, it
is important to remember that the Ni moments included in the scalar
products of the RHS can change in size as well as angle, deviating
from the traditional Heisenberg model.

\begin{figure}
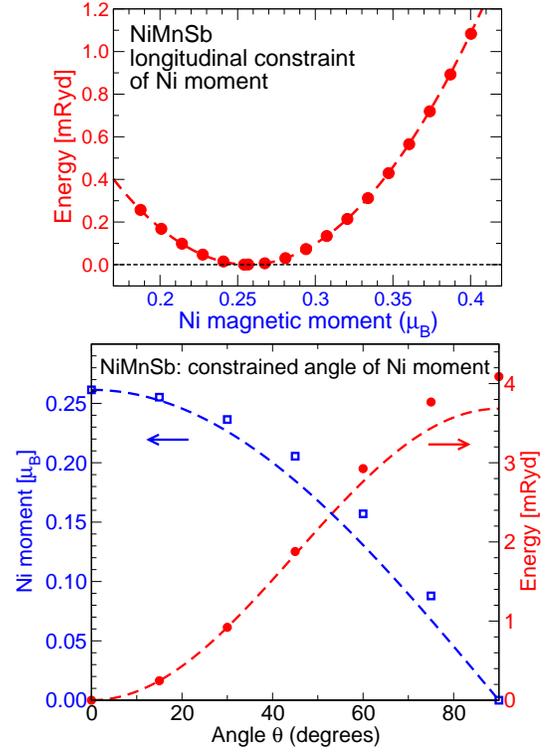

\includegraphics[width=6cm]{fig_energy_long_Bfield.eps}
\includegraphics[width=7cm]{fig_energy_moment_angle_mRyd.eps}
\caption{(color online) Top: 
 Dependence of the total energy on the Ni moment in NiMnSb under the
constraint of an external field acting on the Ni atoms. Circles:
calculated data (within DFT). Dashed curve: parabolic fit to
Eq.~(\ref{eq:L1b}) with $a=53
{\rm mRyd}/\muB^2$.
Bottom:
Dependence of the Ni moment and the total energy on the
  constraining angle $\theta$ tilting the Ni moment away from the Mn
  moment in NiMnSb. Blue squares: $\mu_{\rm eq}(\theta)$ (scale displayed on the
  left-side ordinate). Red circles: $E_{\rm DFT}(\theta)-E_{\rm
    DFT}(0)$ (scale displayed on the right-side ordinate). The broken
  lines correspond to fits to Eqs.~(\ref{eq:L5a}) and
  (\ref{eq:L5b}). In the energy fit, the last part of
  Eq.~(\ref{eq:L5b}) was used with the prefactor $a=55.3 {\rm mRyd}/\muB^2$
  fitted to the low-angle contributions ($\theta=15^\circ$ and
  $30^\circ$). 
  \label{fig:energy_moment_angle}}
\end{figure}

We can derive the parameters $\kappa$ and $a$ from DFT calculations as
follows. From Eq.~(\ref{eq:L1b}) it follows that the equilibrium value
of the Ni moment at $T=0$ and for zero external field is linearly
dependent to the polarizing neigboring Mn moments:
\begin{equation}
\mu_{\rm eq} = \frac{\kappa}{2a} N_c\, M
\label{eq:L4}
\end{equation}
where $N_c=4$ is the coordination number of a Ni atom with respect to
Mn neighbors. From this expression one can deduce the ratio $\kappa/a$
once $M$ and $\mu_{\rm eq}$ have been calculated in the ferromagnetic
state by an {\it ab-initio} method. On the other hand, $a$ can be determined
by introducing into the DFT calculation a constraint on the Ni moment,
either in the form of a longitudinal magnetic field constraining the
magnitude or in the form of a transverse magnetic field
constraining the angle $\theta$ of $\vc{\mu}$ with respect to the moment
directions at the Mn neighbors, but allowing the magnitude to relax to
$\mu_{\rm eq}(\theta)$. The former method results in a parabolic
(for $b=0$) moment-dependence of the total energy, as suggested by
Eqs.~(\ref{eq:L1},\ref{eq:L1b}). The latter method results in
the following dependence, as can be easily found by an energy
minimization of Eq.~(\ref{eq:L1b}) at a given $\theta$:
\begin{eqnarray}
\mu_{\rm eq}(\theta) &=& \mu_{\rm eq}(0) \cos\theta \label{eq:L5a}\\
\Eloc(\theta)-\Eloc(0) &=& a (\vc{\mu}_{\rm eq}(\theta) -\vc{\mu}_{\rm
  eq}(0))^2\nonumber\\
&=&  a \mu_{\rm eq}(0)^2 \sin^2\theta.
\label{eq:L5b}
\end{eqnarray}
If the extended Heisenberg model provides a good approximation to the
energetics of the system, then Eqs.~(\ref{eq:L1b}) and (\ref{eq:L5b})
should both yield a good approximation to the calculated DFT energy
difference, when a constraint is applied on the Ni moment, and the
extracted parameter $a$ for the two cases should be approximately the
same. We find that this is the case in NiMnSb (see
Fig~\ref{fig:energy_moment_angle}).\cite{footnote1} This allows us to
extract the value of $a$ from the DFT calculation and from this the
value of $\kappa$ (equivalent to $J^b_{\Mn-\Ni}$) via
Eq.~(\ref{eq:L4}). Thus the ingredients of formula~(\ref{eq:L3}) are
accessible.

There remains, however, a correction to be made, connected to a
``renormalization'' of the Mn-Mn exchange constants which have been
calculated within the spin-spiral method described in
Section~\ref{sec:Jij}. To clarify the problem we remind the reader
that, when calculating spin spirals, one normally constrains the
direction of the moments but not their magnitude. For the Ni atoms,
where the rigid-moment approximation is not valid, the magnitude $\mu$
is reduced by the spin-spiral formation due to the canting of the
neighboring Mn moments in different directions. This effect provides
an additional, indirect energy contribution to the spin spiral and to
the Mn-Mn interaction, compared to the case that the Ni-moments
magnitude would have been kept constant. It is the spurious
contribution that we mentioned in the introduction to
Sec.~\ref{Sec:longitudinal}. Let us call this contribution
$J_{\Mn-\Ni-\Mn}$. The calculated Mn-Mn interaction consists thus of
two parts:
\begin{equation}
J_{\Mn-\Mn} = J^b_{\Mn-\Mn} + J_{\Mn-\Ni-\Mn}, \label{eq:L5.5}
\end{equation}
with $J^b_{\Mn-\Mn}$ the sought-after bare interaction entering
Eqs.~(\ref{eq:L2.5}) and (\ref{eq:L3}), while
$J_{\Mn-\Mn}$ is a renormalized interaction that is probed by the
spin-spiral DFT calculation. For more distant Mn atoms, which have no
common Ni neighbor, $J^b_{\Mn-\Mn}$ coincides with $J_{\Mn-\Mn}$.

For the derivation of an expression, e.g. in the case of NiMnSb,
giving $J_{\Mn-\Ni-\Mn}$, consider $N_c(=4)$ Mn atoms as nearest
neighbors of a Ni atom. Then the local-energy expression
(\ref{eq:L1b}) becomes
\begin{eqnarray}
\Eloc&=&-\kappa\sum_{n=1}^{N_c}\vc{M}_n\cdot \vc{\mu} + a \mu^2 \nonumber\\
&=& -\frac{1}{2} \sum_{n,n'=1}^{N_c} \frac{\kappa^2}{2a}
\vc{M}_n\cdot\vc{M}_{n'} - \frac{\kappa^2}{4a}N_c\,M^2
\label{eq:L6}
\end{eqnarray}
Here, $n$ and $n'$ run through the Mn moments. The second step follows
under the condition that, in the DFT calculation, the Ni moment
relaxes to the particular equilibrium value that is dictated by the neighboring
Mn-moment directions, i.e., $\vc{\mu}=\frac{\kappa}{2a}\sum_n
\vc{M}_n$. From Eq.~(\ref{eq:L6}) we obtain the indirect interaction:
\begin{equation}
J_{\Mn-\Ni-\Mn} = -\frac{\delta^2 \Eloc}{\delta \vc{M}_n\delta
  \vc{M}_{n'}} = \frac{\kappa^2}{2a}
\label{eq:L7}
\end{equation}

At this point it is important to note that, whether one
chooses to work with $J_{\Mn-\Mn}$ or $J^b_{\Mn-\Mn}$, depends on the
choice of the degrees of freedom. If only the Mn moments are chosen as
degrees of freedom, then $J_{\Mn-\Mn}$ has to be used. If, however,
the Ni moments are also chosen as independent degrees of freedom, then
$J^b_{\Mn-\Mn}$ has to be used together with $J^b_{\Mn-\Ni} =
\kappa$. In the latter case, according to the prescription leading to
Eqs.~(\ref{eq:L2.5}) and (\ref{eq:L7}), we arrive at the
extended Heisenberg Hamiltonian (\ref{eq:L3}) with the appropriate
bare parameters in the nearest-neighbor coupling.

\subsection{Analytical elimination of weak degrees of freedom at $T>0$\label{sec:exact}}

In the previous subsection we discussed the bare and renormalized
parameters of the model arising from total-energy calculations of the
constrained ground state. In this section we examine the case of
thermodynamic quantities at $T>0$, where it is not {\it a priori}
obvious that the same renormalization is still valid if the weak
moments are allowed to fluctuate.  We find that, in the absence of
fourth order terms in \Eloc\ [$b=0$ in Eq.~(\ref{eq:L1})], also at $T>0$
the weak moments can be eliminated in favour of the same renormalized
parameters as the ones appearing at $T=0$. Our conclusion is based on
an exact analytical integration of the weak-moment part of the
partition function in the case $b=0$.

First we observe that the energy functional (\ref{eq:L1b}) under the
action of $\vc{H}^{\rm eff}$ has the same quadratic form as the one
with $\vc{H}^{\rm eff}=0$, only with the minimum shifted to
$\vc{\mu}_{\rm eq}=\vc{H}^{\rm eff}/(2a)$. This simplifies the
integration of the partition function. To see this, we first transform
the Hamiltonian (\ref{eq:L2}) in such a way that the renormalized
interactions $J_{ij}$ appear explicitly. We start by rewriting Eq.~
(\ref{eq:L2}) as:
\begin{eqnarray}
\mathcal{H}(\{\vc{M}_i;\vc{\mu}_l\}) &=& -\frac{1}{2} \sum_{ij} J^b_{ij} \vc{M}_i
\cdot \vc{M}_j \nonumber\\
&&+ a\sum_l \left(\vc{\mu}_l - \frac{\kappa}{2a}\sum_{n(l)} \vc{M}_n\right)^2 \nonumber\\
&& -\frac{\kappa^2}{4a}\sum_l(\sum_{n(l)}\vc{M}_n)^2
\label{eq:L8}
\end{eqnarray}
The indices $i,j$ run over the Mn atoms, the index $l$ over the Ni
atoms, and $n(l)$ over the Mn neighbours of the $l$-th Ni atom. The
last term is now split in an interatomic contribution and an on-site contribution:
\begin{eqnarray}
-\frac{\kappa^2}{4a}\sum_l(\sum_{n(l)}\vc{M}_n)^2 =\nonumber\\
-\frac{1}{2}\sum_l\sum_{\substack{n(l),n'(l)\\ n\neq n'}}
\frac{\kappa^2}{2a} \vc{M}_n \cdot \vc{M}_{n'} +
\sum_l\sum_{n(l)}\frac{\kappa^2}{2a} M_n^2 \nonumber \\
= 
-\frac{1}{2} \sum_{ij; i\neq j} \frac{\kappa^2}{2a} c(i,j) \vc{M}_i \cdot
\vc{M}_{j} + \sum_l\sum_{n(l)}\frac{\kappa^2}{2a} M_n^2
\label{eq:L9}
\end{eqnarray}
In the last step we have converted the sum $\sum_l\sum_{\substack
  n(l),n'(l)\\n\neq n'}$ to a sum over $i,j$, by introducing a
combinatorial factor $c(i,j)$ that counts how many common Ni neighbors
the $i$-th and $j$-th Mn atoms have. From the structure of NiMnSb it
follows that $c(i,j)=1$, if $i,j$ are nearest-neighbors in the Mn fcc
sublattice (i.e. if the distance between $i$ and $j$ is
$\alat/\sqrt{2}$) and $c(i,j)=0$ otherwise. Actually we have recovered
the quantity $J_{\Mn-\Ni-\Mn}=\kappa^2/(2a)$ of Eqs.~(\ref{eq:L6}) and
(\ref{eq:L7}). Finally, the last term of Eq.~(\ref{eq:L9}) is just a
constant which can be omitted. Using the definition (\ref{eq:L5.5})
and Eq.~(\ref{eq:L7}), we combine the first term of the RHS of
Eq.~(\ref{eq:L8}) with the first term of the RHS of Eq.~(\ref{eq:L9})
to obtain the renormailized $J_{ij}$. Then the Hamiltonian to be used
in the partition function takes the form:
\begin{widetext}
\begin{eqnarray}
\mathcal{H}(\{\vc{M}_i;\vc{\mu}_l\}) &=& -\frac{1}{2} \sum_{ij} J_{ij} \vc{M}_i \cdot \vc{M}_j + a\sum_l
\left(\vc{\mu}_l - \frac{\kappa}{2a}\sum_{n(l)} \vc{M}_n\right)^2\nonumber\\
&=& \mathcal{H}_r + a\sum_l
\left(\vc{\mu}_l - \frac{\kappa}{2a}\sum_{n(l)} \vc{M}_n\right)^2
\label{eq:L10}
\end{eqnarray}
where the Heisenberg Hamiltonian involving only the Mn sublattice with
the renormalized interactions, $\mathcal{H}_r=-\frac{1}{2} \sum_{ij}
J_{ij} \vc{M}_i \cdot \vc{M}_j$, has been introduced for convenience
as part of the full Hamiltonian. Note that there is no renormalization
for the Ni-Mn interactions: either they should be completely omitted
within $\mathcal{H}_r$ that includes only the Mn moments, or they
should be given by Eq.~(\ref{eq:L2.5}) in the generalized model. The
spin-spiral result for $J_{\Mn-\Ni}$ does not enter in our theory.

The partition function is:
\begin{eqnarray}
\mathcal{Z}&=& \int d\Omega_1\cdots d\Omega_N \int d^3\mu_1 \cdots
d^3\mu_N \exp\left\{-\frac{\mathcal{H}(\{\vc{M}_i;\vc{\mu}_l\})}{k_BT}
\right\} \label{eq:L11a}\\
&=&  \int d\Omega_1\cdots d\Omega_N \exp
\left\{-\frac{\mathcal{H}_r}{k_B T} \right\}
\int d^3\mu_1 \cdots d^3\mu_N 
\exp\left\{-\frac{ a\sum_l
(\vc{\mu}_l - \frac{\kappa}{2a}\sum_{n(l)} \vc{M}_n)^2 }{k_B T}\right\}\\
&=&\int d\Omega_1\cdots d\Omega_N 
\exp
\left\{-\frac{\mathcal{H}_r}{k_B T} \right\}
 \left(\frac{\pi k_BT}{a}\right)^{\frac{3N}{2}}\\
&=& \mathcal{Z}_r  \left(\frac{\pi k_BT}{a}\right)^{\frac{3N}{2}} 
\label{eq:L11}
\end{eqnarray}
where $\int d\Omega_1\cdots d\Omega_N$ denotes an integration over the
Mn-moment solid angles. The integration over the Ni-moments $\int
d^3\mu_1 \cdots d^3\mu_N$ contains only the exponential of a complete
square and has been analytically integrated to $(\pi k_BT/a)^{3N/2}$,
i.e., it corresponds to the partition function of $3N$ uncoupled
harmonic degrees of freedom and is independent of the value of
$\vc{M}_n$. $\mathcal{Z}_r$ is the partition function corresponding to
the Hamiltonian $\mathcal{H}_r$. The magnetization $\langle
\vc{M}\rangle$ of the Mn sublattice also turns out to depend only on
the renormalized partition function (and Hamiltonian). The same is
true for any moment-moment correlation function of the Mn sublattice,
$\langle(M_{1x})^{m_{1x}}(M_{1y})^{m_{1y}}\cdots(M_{Nz})^{m_{Nz}}\rangle$,
$\alpha\in (x,y,z)$, where the $m_{n\alpha}\geq0$ are integer
exponents defining the order of the correlation function. To see this
we consider explicitly
\begin{eqnarray}
\langle(M_{1x})^{m_{1x}}(M_{1y})^{m_{1y}}\cdots(M_{Nz})^{m_{Nz}}\rangle &=& \frac{1}{\mathcal{Z}}
 \int d\Omega_1\cdots d\Omega_N
 (M_{1x})^{m_{1x}}\cdots(M_{Nz})^{m_{Nz}}
 \exp\{-\mathcal{H}_r/k_BT\}
\nonumber\\
&&\times \int d^3\mu_1 \cdots d^3\mu_N 
\exp\biggl\{-\frac{ a\sum_l
(\vc{\mu}_l - \frac{\kappa}{2a}\sum_{n(l)} \vc{M}_n)^2}{k_B T} \biggr\} \label{eq:L12a}\\
&=&
\frac{1}{\mathcal{Z}_r}\int d\Omega_1\cdots d\Omega_N  (M_{1x})^{m_{1x}}\cdots(M_{Nz})^{m_{Nz}} \exp\{-\mathcal{H}_r/k_BT\}
\label{eq:L12}
\end{eqnarray}
\end{widetext}
where the Hamiltonian in the exponential has again been split in two
terms according to Eq.~(\ref{eq:L10}) and the integration over
$d^3\mu_1 \cdots d^3\mu_N$ has again been 
carried out analytically, cancelling out the factor $(\pi k_BT/a)^{3N/2}$ in
the full partition function. But expression (\ref{eq:L12}) is just the expression for the
correlation function of only
the Mn sublattice with the renormalized Heisenberg interactions. Thus we see
that, as far as the Mn-sublattice magnetization is concerned, one can
use the exchange interactions calculated by the spin-spiral method
that are renormalized by construction,
while neglecting the Ni moments and the Mn-Ni interaction. If the Ni
moments are to be included, then the bare interactions have to be used
in an extended model, however, the Mn-sublattice properties in the
two cases will be the same. Note, finally, that this exact
result is based on the fact that the exponent in the integration over
$d^3\mu_1 \cdots d^3\mu_N$ contains a complete square, i.e., that the
``harmonic'' approximation, $b=0$, is valid; if $b\neq 0$ and one
proceeds to an elimination of the weak degrees of freedom, then the
resulting renormalized strong-moment Hamiltonian can have higher-order
terms (biquadratic, four-spin, etc.) with temperature-dependent
parameters. 

A semi-analytical result follows in an analogous way also for the
average Ni moment and fluctuation amplitude per atom:
\begin{eqnarray}
\langle\vc{\mu}\rangle &=& \frac{\kappa}{2a} N_c\, \langle \vc{M}\rangle
\label{eq:L13} \\
\langle\mu^2\rangle &:=& \frac{1}{N} \langle [\sum_{l=1}^N
\vc{\mu}_l]\rangle^2\label{eq:L14} \\
&=&\frac{1}{a}\frac{3}{2}k_BT + \left(\frac{\kappa}{2a}N_c\right)^2 \frac{1}{N} \langle [\sum_{i=1}^N
\vc{M}_i]^2\rangle \nonumber\\ 
&=& \frac{1}{a}\frac{3}{2}k_BT + \left(\frac{\kappa}{2a}N_c\right)^2 \langle M^2 \rangle
\label{eq:L15}
\end{eqnarray} 
I.e., there is a ``harmonic'' part and a part induced by the
fluctuation of the Mn sublattice. The former is independent of the
number of atoms, while the latter, $\langle M^2 \rangle := \langle
[\sum_{i=1}^N \vc{M}_i]^2\rangle/N$, increases proportionally to the
number of magnetic atoms in the system for $T<\Tc$, as in normal
Heisenberg systems.\cite{footnoteHMK} The longitudinal susceptibility
$\chi_{\Ni}$ of the Ni sublattice can be found in a completely
analogous way:
\begin{eqnarray}
k_BT\,\chi_{\Ni} &=& \langle \mu_z^2\rangle - \langle \mu_z \rangle^2
\nonumber\\
&=& \frac{1}{a}\frac{k_BT}{2} + \left(\frac{\kappa}{2a}N_c\right)^2
(\langle M_z^2\rangle -\langle M_z\rangle^2) \label{eq:L16a}\\
&=& k_BT \left[ \frac{1}{2a} + \left(\frac{\kappa}{2a}N_c\right)^2
  \chi_{\Mn}\right]
\label{eq:L16}
\end{eqnarray}
i.e., there is again a harmonic part and a part proportional to the Mn
sublattice susceptibility. (In Eq.~(\ref{eq:L16a}), $z$ is implied to
be the direction of magnetization.)

According to these results, as long as the approximation $b=0$ holds,
one can deduce the thermodynamic properties of the weak sublattice by
merely a calculation on the strong sublattice, avoiding the extra
numetical cost that a full Monte Carlo simulation entails. Actually
this procedure can be carried out for higher-order correlation
functions of the Ni moments,
$\langle(\mu_{1x})^{m_{1x}}(\mu_{1y})^{m_{1y}}\cdots(\mu_{Nz})^{m_{Nz}}\rangle$
($\alpha\in (x,y,z)$), where once again the $m_{l\alpha}\geq0$ are
integer exponents defining the order of the correlation function. In
the resulting formula the integration over $d^3\mu_l$ can be carried
out analytically, yielding a sum of correlation functions of the
$\vc{M}_i$ of the form:
\begin{widetext}
\begin{eqnarray}
\langle\mu_{1x}^{m_{1x}}\cdots\mu_{Nz}^{m_{Nz}}\rangle &=&
\frac{1}{\mathcal{Z}}\int d\Omega_1\cdots d\Omega_N
e^{-\frac{\mathcal{H}_r}{k_BT}} \int d^3\mu_1\cdots d^3\mu_N 
(\mu_{1x})^{m_{1x}}\cdots(\mu_{Nz})^{m_{Nz}}
\exp\biggl\{-\frac{ a\sum_l
(\vc{\mu}_l - \frac{\kappa}{2a}\sum_{n(l)} \vc{M}_n)^2}{k_B T} \biggr\}
\nonumber\\
&=&
\frac{1}{{\mathcal{Z}_r}} \int d\Omega_1\cdots d\Omega_N
e^{-\frac{\mathcal{H}_r}{k_BT}} \prod_{l=1}^N\prod_{a\in xyz}
\sum_{\substack{p=0\\p\ \rm
    even}}^{m_{l\alpha}}
\binom{m_{l\alpha}}{p}(p-1)!!\left(\frac{k_BT}{2a}\right)^{p/2}\Bigl(\frac{\kappa}{2a}\sum_{n(l)}M_{n\alpha}\Bigr)^{m_l(\alpha)-p}
\label{eq:L18}
\end{eqnarray}
\end{widetext}
where a change of variables $\mu_{l\alpha}\rightarrow
\mu_{l\alpha}+\frac{\kappa}{2a}\sum_{n(l)}M_{n\alpha}$ has been
performed, the binomial expansion of
$(\mu_{l\alpha}+\frac{\kappa}{2a}\sum_{n(l)}M_{n\alpha})^{m_{l\alpha}}$
has been used and we have taken into account that integrals of the
type $\int_{-\infty}^{\infty}x^p
e^{-ax^2}=\sqrt{\frac{\pi}{a}}(p-1)!!/(2a)^{p/2}$ for even $p$ and
vanish for odd $p$ [for the $p=0$ term we accept the convention
$(-1)!!=1$]. Due to the presence of products of $M_{n\alpha}$ in
Eq.~(\ref{eq:L18}) it is clear that this expression reduces to a sum
of correlation functions of the $\vc{M}_i$ within the Hamiltonian
$\mathcal{H}_r$. The summations in this expression are too involved to
arrive at a general closed form, however, Eqs.~(\ref{eq:L13}),
(\ref{eq:L15}) and (\ref{eq:L16}) are special cases of application of
this formula. Obviously, mixed correlation functions
between strong and weak moments can also be reduced to strong-moment
correlation functions of the renormalized Hamiltonian by first
eliminating the weak moments following the same prescription.

We should stress that choosing to eliminate the weak moments in favour
of renormalized interactions does not mean that the weak moments are
physically less valid as degrees of freedom; to do so is merely a
matter of mathematical or computational convenience, especially since
efficient methods exist for the calculation of thermodynamic
quantities within the classical Heisenberg model.

\subsection{Calculations in NiMnSb\label{sec:NiMnSb}}

We exemplify the above results with calculations in NiMnSb. First we
establish which interactions can be neglected. To this end we
performed calculations of the exchange coupling parameters by the
spin-spiral method. As it turns out, the Ni-Ni interactions are
negligible. About the Mn-Mn interactions, Fig.~\ref{fig:exchange}b
suggests that it should be enough to include up to 2nd neigbors, but
this is misleading; we find by Monte Carlo calculations a 2nd neighbor
approximation leads to an overestimation of approximately 100~K in
\Tc\ compared to the value including also more distant neighbors,
therefore we include Mn-Mn interactions up to a distance of three
lattice parameters.  Among the Mn-Ni interactions
(Fig.~\ref{fig:exchange}b) only the nearest-neighbor coupling has some
small influence, changing \Tc\ by a modest 20~K. Given the conclusions
of the previous subsection, the Mn-Ni interaction should be excluded
for an estimation of \Tc, since the Mn-Mn interactions are already
renormalized (in the corresponding spin-spiral calculations the
magnitude of the Ni moment was allowed to relax). The Mn-Ni
interaction must be included if one wishes to extract information on
the behavior of the Ni magnetization; then, however, the interaction
type and strength has to be corrected, since the Ni atoms belong to a
soft-magnetic sublattice.

After deriving the exchange parameters $J_{ij}$ by the spin-spiral
method, utilizing the force theorem which according to
Fig.~\ref{fig:interstitial}a is accurate enough for this purpose, we
performed constrained-angle calculations for the Ni moments with full
self-consistency. From the total-energy results shown in
Fig.~\ref{fig:energy_moment_angle}, together with Eq.~(\ref{eq:L5b}),
we deduce a value of $a=55.3~{\rm mRyd}/\muB^2$. Given this, together
with the ground-state magnetic moments of Mn $M=3.71~\muB$ and of Ni
$\mu_{\rm eq} = 0.26~\muB$, Eqs.~(\ref{eq:L2.5}) and (\ref{eq:L4})
yield a value of $J^b_{\Mn-\Ni}\equiv \kappa=1.94~{\rm
  mRyd}/\muB^2$. This bare value $J^b_{\Mn-\Ni}$ is significantly
different than the value of $0.92~{\rm mRyd}/\muB^2$ for the Mn-Ni
interaction, which was derived from the spin-spiral calculation
(Fig.~\ref{fig:exchange}B).  Finally, from Eqs.~(\ref{eq:L5.5}) and
(\ref{eq:L7}) we can extract the bare value of the nearest-neighbor
Mn-Mn interaction: $J_{\Mn-\Mn}^b = 0.066~{\rm mRyd}/\muB^2$, which
shows a reduction of about $1/3$ compared to the corresponding
renormalized value. Note that the equilibrium moments have to be
multiplied to these values if comparison is to be made with the
energies shown in Fig.~\ref{fig:exchange}B. Then one obtains
$J^b_{\Mn-\Mn}M^2=0.91~{\rm mRyd}$ and $J^b_{\Mn-\Ni}M\mu_{\rm
  eq}=1.87~{\rm mRyd}$; i.e., the Mn-Mn bare interaction is weaker
than the Mn-Ni, which is not surprising, since the Ni polarization
stems from a direct, nearest-neighbor exchange interaction with Mn.

\begin{figure}[ht]
\begin{center}
\includegraphics[width=7.5cm]{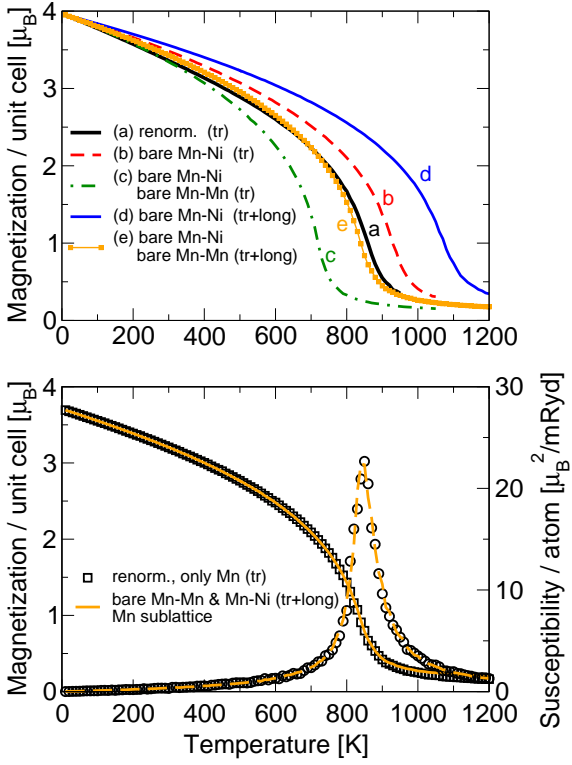}
\caption{(color online) Top: Monte Carlo results on the magnetization of NiMnSb under
  various assumptions on the model and the interactions. (a-c): Only
  transverse fluctuations are taken. (a): Interactions as derived by
  force-theorem, spin-spiral calculations, including Mn-Mn and
  Mn-Ni. (b) Same as (a), but with bare Mn-Ni interactions
  $J^b_{\Mn-\Ni}$ (Eq.~\ref{eq:L2}). (c) Same as (b), but additionally
  with bare Mn-Mn interactions $J^b_{\Mn-\Mn}$ (given by
  Eqs.~(\ref{eq:L5.5},\ref{eq:L7})). (d-e): Also longitudinal
  fluctuations of the Ni-moment are allowed. (d) Bare Mn-Ni
  interactions. (e) Bare Mn-Ni and Mn-Mn interactions. Evidently,
  taking the Mn-Ni bare interaction alone leads to an overestimation
  of the excitation energies and \Tc, which is corrected when the bare
  Mn-Mn interaction is also taken. Bottom: Magnetization curve and
  susceptibility of the Mn sublattice in NiMnSb calculated within
  different models. Squares (magnetization) and circles
  (susceptibility): Only transverse fluctuations with renormalized
  Mn-Mn interactions, neglecting completely the Ni moments. Full line:
  Transverse and longitudinal fluctuations are allowed for the Ni
  atoms, only transverse for Mn, with bare Mn-Mn and Mn-Ni parameters;
  only the Mn sublattice magnetization and susceptibility is shown
  (i.e., corresponding to the Mn part of curve (e) in the top
  panel). Note on the susceptibility units that $1\muB^2/{\rm mRyd} =
  4.254\times 10^{-3}\muB/{\rm Tesla}$. The temperature dependence of
  the magnetization and susceptibility is identical in the two cases
  for reasons that are explained in the text. \label{fig:renorm1}}
\end{center}
\end{figure}

\begin{figure}[ht!]
\begin{center}
\includegraphics[width=7.5cm]{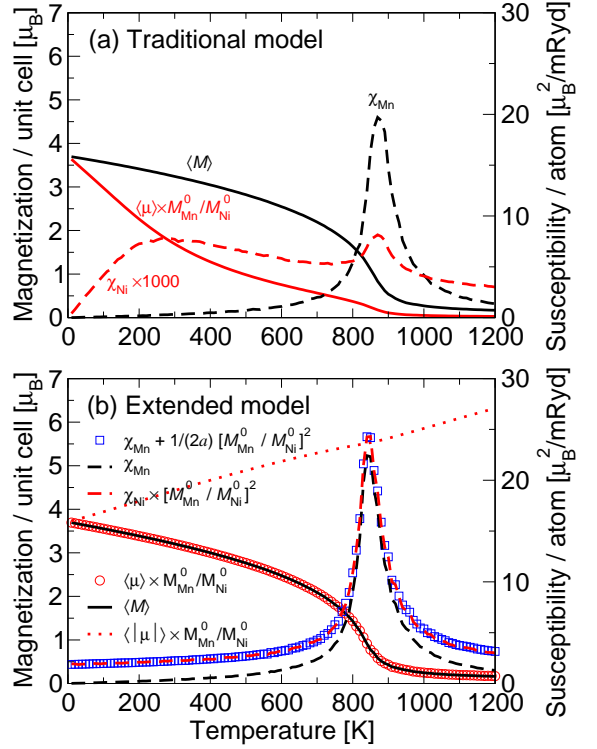}
\caption{(color online) Magnetization curves and sublattice susceptibility in the
  traditional and extended model. The notation is as follows: $\langle
  M\rangle$ and $\langle \mu\rangle$ stand for the
  temperature-dependent magnetization of the Mn and Ni sublattices,
  respectively; $\langle |\mu|\rangle$ stands for the thermal average
  of the absolute value of the Ni local moment; $\chi_{\rm Mn,Ni}$ for
  the sublattice susceptibilities of Mn and Ni; and $M_{\rm Mn,Ni}^0$
  for the ground-state local moments.  (a) Traditional model (only
  transverse fluctuations of the moment) with the interactions derived
  by force-theorem from spin-spiral calculations. The Mn magnetization
  shows a typical ferromagnetic behavior with a susceptibility peak at
  \Tc. On the other hand the Ni magnetization drops rather fast, with
  an atypical susceptibility showing a plateau over a wide region. (b)
  Extended model (longitudinal fluctuations of the Ni moment are
  allowed) with bare Mn-Ni and Mn-Mn interactions. The Mn
  magnetization hardly changes compared to (a), and the \Tc\ is very
  similar, but the behavior of the Ni magnetization and susceptibility
  are completely different, following the critical behavior of the Mn
  sublattice. At $T=0$, the Ni susceptibility does not vanish,
  behaving as it is expected for a system with non-rigid moments,
  reaching the value $\chi=1/(2a)=0.009~\muB^2/{\rm mRyd}$ at
  $T=0$. The average of the magnitude of the local Ni moment increases
  with temperature. The Monte Carlo simulations included 2048 unit
  cells (4096 atoms).  
The
  Ni-related quantities have been scaled up by the annotated factors
  in order to demonstrate the agreement with Eqs.~(\ref{eq:L13}) and
  (\ref{eq:L16}): $\langle \mu\rangle$ and $\langle |\mu|\rangle$ by a
  factor $M_{\rm Mn}^0/M_{\rm Ni}^0=14.2$ and $\chi_{\rm Ni}$ by
  $(M_{\rm Mn}^0/M_{\rm Ni}^0)^2=201$.  In particular the blue squares
  show the validity of Eq.~(\ref{eq:L16}), i.e., the derivation of the
  Ni from the Mn susceptibility. The latter coincides with the
  $\chi_{\Mn}$ of the traditional renormalized model, as is
  demonstrated in Fig.~\ref{fig:renorm1} (bottom
  panel). \label{fig:renorm2}}
\end{center}
\end{figure}

Next we present a series of Monte-Carlo-simulation results, examining
the effect and importance of the bare interactions. We performed
simulations in the framework of the traditional model (only transverse
fluctuations allowed) as well as the extended model (longitudinal
fluctuations also allowed), with the renormalized and bare parameters.
\new{Putting it all together, we must substitute the above-found
  parameters $J_{\Mn-\Mn}^b$, $J_{\Mn-\Ni}^b$, and $a$, into the
  extended Hamiltonian (\ref{eq:L3}) which is to be treated with a
  Monte Carlo method where the (weak) Ni moments $\vc{\mu}_l$ should
  be allowed to vary in length and direction while the (strong) Mn
  moments should be allowed to vary in direction only. In this
  particular example, a distinction between bare and renormalized
  interactions was made only for the nearest-neighbour Mn-Mn and Mn-Ni
  coupling. The longer-range Mn-Ni coupling was assumed to vanish thus
  there can be no renormalization in the longer-range Mn-Mn coupling
  [as the distant Mn atoms have no common Ni neighbor, i.e.,
  $J_{\Mn-\Ni-\Mn}=0$ for distant atoms in Eq.~\ref{eq:L5.5}]. The Ni-Ni coupling was
  also assumed to vanish. The results are to be compared with
  calculations employing the traditional Hamiltonian
\begin{equation}
\mathcal{H}(\{\vc{M}_i\}) = -\frac{1}{2}\sum_{ij \in \Mn} J_{ij}
\vc{M}_i\cdot\vc{M}_j 
\end{equation}
containing only the Mn moments that vary in direction, where now
$J_{ij}$ are the Mn-Mn renormalized parameters, i.e., obtained from
the spin-spiral method presented in Sec.~\ref{sec:Jij} and shown
(multiplied by the moments) in Fig.~\ref{fig:exchange}b.} In both cases
the Mn-Mn interaction was included up to a distance of three lattice
parameters. The first set of results is contained in
Fig.~\ref{fig:renorm1}, where magnetization curves are shown,
calculated within different assumptions. Here we omit showing the
susceptibiliy, but the Curie temperature can be recognized by the
characteristic inflection point of the magnetization curve. The
central results are included in curves (a) and (e) of the top panel,
as well as in the bottom panel, while curves (b), (c) and (d) merely
show that neglecting some of the degrees of freedom or some of the
bare constants leads to almost arbitrary results. In the simulations
including longitudinal moment fluctuations, more Monte Carlo steps
were necessary, typically by an order of magnitude, in order to arrive
at the same quality as in the simulations including only transverse
fluctuations.

Curve (a) shows the result of the traditional model with
spin-spiral-derived (i.e., renormalized) Mn-Mn exchange parameters,
including spin-spiral-derived Mn-Ni interactions. This results in
$\Tc=870~{\rm K}$. Curve (b) includes the bare Mn-Ni parameters, but
stays within the traditional model. \Tc\ increases by a modest amount
of 60~K, since the bare Mn-Ni parameters are stronger than the
spin-spiral-calculated ones. Curve (c) also stays within the
traditional model, but includes the bare Mn-Mn interaction, which is
weaker by a factor $1/3$ compared to the renormalized value; \Tc\
drops significantly to 730~K. This coincides with the experimental
value, but the coincidence is probably fortuitous: the longitudinal
fluctuations at Ni, that are essential to the derivation of the bare
parameters, are yet unaccounted for in the simulation (see, however,
the discussion on the phase space measure in
subsection \ref{sec:remarks}). Next comes curve (d), where the longitudinal fluctuations
are allowed for in the simulation, but taking the bare Mn-Ni and the
renormalized Mn-Mn exchange. The increase in \Tc\ (1090~K) with
respect to curve (b) (same parameters but rigid Ni moment) is
striking. The difference stems from the fact that the Ni local moment
becomes larger at high temperatures (see discussion on
Fig.~\ref{fig:renorm2} below), thus the ferromagnetic contribution
$J_{\Mn-\Ni}\vc{M}\cdot\vc{\mu}$ to the Hamiltonian is effectively
strengthened. Finally, if we account also for the (weaker) bare Mn-Mn
coupling in the extended model, we obtain curve (e), which is very
close to the original curve (a), also with a very similar
$\Tc=850$~K. This value, however, is the same that one obtains if the
traditional model is used, but with the Ni moments completely
neglected. In fact, the corresponding magnetization curve falls
exactly on top of the Mn-contribution to curve (e), and the same is
true for the Mn-sublattice susceptibility.  This striking
agreement is also demonstrated in the bottom panel of
Fig.~\ref{fig:renorm1}, and is expected on the basis of
Eq.~(\ref{eq:L12}) and the discussion thereafter.

Let us consider now the contrast between the traditional model with
spin-spiral calculated interactions and the extended model with bare
interactions. Monte-Carlo results on these are shown in
Fig.~\ref{fig:renorm2}a and b, representing the traditional and
extended model respectively. The main difference lies in the behavior
of the Ni-sublattice magnetization $\langle\mu_z\rangle$ and susceptibility
$\chi_{\Ni}$. For better comparison we have scaled up these
quantities. In Fig.~\ref{fig:renorm2}a we see that the traditional
model results in a Ni magnetization that drops rather fast with
temperature, much faster than the Mn magnetization. This is due to the
relatively weak coupling between the Mn and Ni sublattices. Although
the difference in energy scale is not directly obvious from
Fig.~\ref{fig:exchange}, recall that a Mn atom is surrounded by 4 only
Ni atoms but by 12 Mn atoms at distance $\alat/\sqrt{2}$ and 6 Mn atoms at
distance $\alat$, where the exchange coupling is still appreciable. The Ni
susceptibility, $\chi_{\Ni}$ shows a large plateau below \Tc\ and
follows the critical peak of Mn at \Tc\ rather weakly. The behavior of
the Mn magnetization and susceptibility, on the other hand, is
characteristic of a ferromagnetic phase transition.

In Fig.~\ref{fig:renorm2}b we see results within the extended
model. Here, the Ni magnetization follows the behavior of Mn. The
susceptibility $\chi_{\Ni}$ starts off at a finite value at $T=0$,
which coincides with the value given by Eq.~\ref{eq:L1c}, in contrast
to the vanishing susceptibility at $T=0$ for rigid-moment systems. At
\Tc\ the Ni susceptibility shows a peak following the critical
behavior. The Ni magnetization and susceptibility have been scaled up
by appropriate factors to show that Eqs.(\ref{eq:L13}) and
(\ref{eq:L16}) are reproduced by the Monte Carlo simulation. A
technical point to be mentioned for accuracy is that, in calculating
$\langle\mu_z\rangle$ and $\langle\mu_z^2\rangle$ in the simulation
one should take the projection of $\sum_l\vc{\mu}_l$ in the direction
of the total moment at each Monte Carlo step, instead of making the
approximation to calculate the average and variance of
$|\sum_l\vc{\mu}_l|$.

It is also interesting to see that the average magnitude of
the local Ni moment (dotted line) increases monotonically with
temperature. This effect, pointed out by
Sandratskii\cite{Sandratskii08} in an analytical low-temperature
approximation for NiMnSb, is connected to the fact that even above
\Tc\ there is some short-range order in the system, so that the
equilibrium value of
$\langle\vc{\mu}\rangle=\frac{\kappa}{2a}N_c\langle\vc{M}\rangle$ is
non-vanishing. On top of this, the fluctuations of $\vc{\mu}$ shift
the value of
$\langle|\vc{\mu}|\rangle=\langle(\mu_x^2+\mu_y^2+\mu_z^2)^{1/2}\rangle$
to higher values. The increase is expected to cease when $\mu$ reaches
such high values that the approximation $b=0$ is no more valid (so
that the fluctuations are moderated by the fourth-order term); this
correction does not apply for NiMnSb, however, at least at
temperatures as high as \Tc, since the constrained-DFT calculation
shown in Fig.~\ref{fig:energy_moment_angle} (top panel) yields a
quadratic energy dependence at values of $\mu$ comparable to the ones
close to \Tc\ in Fig.~\ref{fig:renorm2}b.

Experimentally, the temperature dependent magnetization of NiMnSb
follows Bloch's $T^{3/2}$ law up to about
70-100~K,\cite{Hordequin96,Ritchie03} which cannot be reproduced
within classical Heisenberg models.  Element (or sublattice) specific
experiments were done using neutron scattering\cite{Hordequin97} and
X-ray magnetic circular dichroism (XMCD).\cite{Borca01} Neutron
scattering \cite{Hordequin97} at 15~K and 260~K shows a thermally
stable Ni moment $(0.18\ \mu_B)$, but a decreasing Mn moment (dropping
from $3.79\ \mu_B$ at 15~K to $3.55\ \mu_B$ at 260~K). Contrary to
this, according to the XMCD results of Ref.~[\onlinecite{Borca01}]
\emph{both} the Ni and Mn moments drop rapidly at 80~K to half their
ground-state values, and then level off up to at least 250~K. Such
behavior would be counter-intuitive; the authors in
[\onlinecite{Borca01}] write that surface effects possibly complicate
the interpretation of MCD data.

\section{Final remarks\label{sec:remarks}}

\subsection{Remarks on the treatment of the weak moments and on the
  concept of renormalization}

Recently, Wysocki, Glasbrenner and Belashchenko\cite{Wysocki08} (WGB)
presented a study of a classical spin-fluctuation model. Their model
Hamiltonian is similar to the one that we use here, with the
difference that in the WGB paper all atoms can change the magnitude of
their moments, the 4th-order term is not neglected, and in practice
only one atomic species is considered in their calculations. WGB point
out that the magnetic moments are not canonical variables, therefore
there is no obvious way how to choose the phase space measure, which
should therefore be given as part of the model together with the
Hamiltonian. In the present work we have chosen what they call uniform
phase space measure, which basically amounts to dividing the
$\vc{\mu}$-space in equal-volume infinitesimal cells with equal
integration weight, and which is the most common choice in the
literature.\cite{footnote2} It also amounts to taking a simple
$d^3\mu$ integration with no further weight in
Eqs.~(\ref{eq:L11a},\ref{eq:L12a},\ref{eq:L18}). Different choices of measure can lead to
qualitatively different results, e.g. a fast drop of magnetization in
the weakly-magnetic sublattice, as also shown by
Sandratskii,\cite{Sandratskii08} or even a first-order transition,
as WGB find.\cite{Wysocki08} The correct measure can only be certified
by the best classical approximation to the full quantum-mechanical
solution, which, however, remains an open problem. However, if it is
assumed that the correct measure is not uniform but e.g. proportional
to $\mu^{-2}$ (this was one alternative choise by WGB), then the
weak-sublattice magnetization will drop fast at low temperatures, so
that close to \Tc\ only the bare parameters of the strong sublattice
would be relevant.

\new{ Mryasov et al.\cite{Mryasov05} have discussed the idea of
  renormalized interactions in the case of FePt alloys.  Related is
  also the work by Polesya {\it et al.}\cite{Polesya10} who adopt a
  model for FePd and CoPt alloys. In these works, the weak moments of
  Pd or Pt are determined from the strong moments of their neighbors
  via the susceptibility. However, even at higher temperatures the
  weak moments are not treated as independent variables that can
  fluctuate (either in direction or in length) but rather as enslaved
  quantities to their immediate neighbourhood i.e., their role in the
  thermodynamics is only to mediate an additional interaction between
  the strong moments.  Our idea of renormalization has an analogous
  starting point but goes along a different path, since we consider
  the weak moments in the spin Hamiltonian as independent fluctuating
  variables.

  In addition, the main novelty of the present study in this respect
  is the formal and numerical proof of the equivalence of two
  approaches: all thermodynamic quantities can be derived by using
  either the extended model with bare interactions or the traditional
  model with renormalized interactions. Furthermore it is shown in the
  present study that the renormalized intercations are actually the
  ones that are harvested by the spin-spiral approach within
  density-functional calculations with no further manipulations;
  however, manipulations are necessary if one wishes to extract the
  bare parameters. The present result holds} under the assumption of a
uniform phase space measure and of a quadratic on-site energy. The
latter seems to hold true e.g. for NiMnSb and for
FePt,\cite{Mryasov05,Mryasov05b,Sandratskii08} but not, for example,
in the case of FeRh\cite{Mryasov05b,Sandratskii11} where higher-order
corrections are necessary. \new{In case that either of these two
  requirements is not met, most probably a hypothesis of
  temperature-independent renormalized interactions cannot be
  justified on the grounds of fluctuating weak moments, but should
  instead be conjectured as an ad-hoc hypothesis within the model.}

Bruno\cite{Bruno03} also introduces a concept of renormalized exchange
parameters. However, his approach encapsulates different physics than
our present approach. Bruno's renormalization corrects for a
systematic error, mainly due to the difference of the
constraining-field direction to the resulting moment direction when
the force theorem is applied. Our renormalization on the other hand
concerns the error due to the reduction of the weak-moment magnitude
when the strong moments are tilted.

Yet another concept of renormalization is described by Lounis and
Dederichs.\cite{Lounis10} Using a multiple-scattering approach, they
consider the energy expansion as a function of the angles between
moments. As they find, at high angles corrections are necessary to the
phenomenological Hamiltonian (e.g., biquadratic or four-spin
terms). However, at low angles these corrections can be partly
included in the Heisenberg model via a renormalization of the exchange
parameters, recovering correct energy scales and Curie temperatures.

\subsection{Remarks on the prediction of the Curie temperature\label{sec:remarksII}}

The Curie temperatures calculated within the mainstream approach to
the adiabatic spin-dynamics are in many cases in agreement with
experiment to within 10-15\%, but with no obvious systematics toward
over- or underestimation.  The main source of error is not
clear. Considering the most serious approximations made, error can
stem from
\begin{itemize}
\item
(i) the use of local density functional theory (LDA or GGA)
for total energy calculations without further corrections for electron
exchange and correlation. 
\item
(ii) the use of the adiabatic approximation,
\item
(iii) the assumption of a classical, rather than quantum, Heisenberg
model,
\item
(iv) the assumption that the exchange constants do not change as a
function of temperature, 
\item
(v) the assumption of rigid spins of the strong moments in the
Heisenberg model.
\end{itemize}
In general, these factors have possibly different weight for different
materials. Concerning point (i), theories that provide a better
treatment of correlations exist, e.g., the LSDA+$U$ or LSDA combined
with dynamical mean field theory at zero or finite
temperatures. Also within such theories exchange parameters can be 
derived (see, e.g., Ref.~\onlinecite{Katsnelson00}). However,
parameters are required (as is the Coulomb repulsion $U$), and it is
usually not obvious how to determine these uniquely.

As for point (ii), there are promising developments in the calculation
of magnon spectra (including magnon lifetime effects) within
time-dependent density functional theory~\cite{Buczek:09,Buczek:11,
  Rousseau:12} or many-body perturbation
theory~\cite{Sasioglu:10}. These can prove very useful in the future
(they can also be extended to finite temperatures), however at this
point they are computationally too demanding for systematic
calculations of the Curie temperature.

Point (iii) (the classical assumption), can be improved upon
by using the random phase approximation to solve the quantum
Heisenberg model. However, for itinerant electron systems the local
moment $M$ does not corespond to some integer or half-integer value of
the spin $S$ either in the form $M=\sqrt{S(S+1)}\ \muB$ or in the form $M_z=S\ \muB$. In fact,
calculations of Heusler compounds in Ref.~[\onlinecite{Sasioglu05}]
have shown that for a reasonable choice of $S$ the Curie temperature
is strongly overestimated, while the classical limit of the random
phase approximation, with a choice of
large $S$ (with appropriate normalization of the exchange parameters
so that the product $J_{ij}S_iS_j$ remains constant), results in reasonable values
of \Tc. Therefore, a quantum Heisenberg model is perhaps a better for
a correct description of the shape of the magnetization curve $M(T)$,
but a poor choice for a correct \Tc, at least in itinerant electron
systems, if the exchange constants are calculated within the adiabatic
approximation. This conclusion is in accordance to the spirit of
adiabatic spin dynamcics,\cite{Antropov96} where the effective
interactions $J_{nn'}$ correspond to the equation of motion of the
expectation value of the local moments, i.e., to classical quantities,
not operators.

Corrections to point (iv) can be treated within local density
functional theory if the exchange constants are calculated starting
from a disordered local moment state. This requires use of the
coherent-potential approximation (CPA), and has been proposed for
example in Ref.~[\onlinecite{Shallcross05}]. The use of the CPA for
the description of the disordered local moment state at \Tc\
underestimates the existence of magnetic short range order (which is
known to be present); however, it constitutes a promissing approach,
since it can be systematically improved e.g. by the use of a non-local
CPA.\cite{Rowlands03}

Finally, point (v) becomes a serious approximation in systems of weak
magnetic moments, such as ferromagnetic Ni, and has been widely
discussed in the literature as we noted in
Sec.~\ref{Sec:longitudinal}. Corresponding corrections for
multiple-scattering based methods have been recently proposed, e.g.,
by Bruno\cite{Bruno03} and Shallcross and
co-workers.\cite{Shallcross05} In a more recent work by Ruban {\it et
  al.},\cite{Ruban07} based on an expansion of the energy within the
disordered local moment state, promising results were obtained showing
the fundamental importance of longitudinal corrections to the local
moment for \Tc\ in ferromagnetic Ni.

\section{Summary and conclusions\label{Sec:summary}}

\new{In the first part of this work we have investigated the calculation of
interatomic exchange constants that we implemented in the FLAPW method
based on the concept of adiabatic spin dynamics.}  The exchange
constants are harvested by an inverse Fourier transformation involving
static spin-spiral energies.\new{\cite{Halilov98}} Symmetry relations
obeyed by the spin spiral energies have been found to greatly reduce
the numerical effort, in particular regarding confinement of the
inverse Fourier transformation in the irreducible wedge of the
Brillouin zone. Furthermore, the force-theorem approximation has been
tested and found to be adequate for small cone angles of the spin
spirals. However, we have shown that application of the force theorem
requires special treatment of the intersitial region, namely setting
there the magnetic part of the exchange-correlation field to zero.

In the second part of the present work we have proposed a way to
explore multicomponent systems where a magnetically strong sublattice
coexists with a magnetically weak sublattice, necessiating a
consideration of \new{longitudinal and transverse changes of the weak
  local magnetic moments while they are still treated as independent
  variables.} We find the rigorous result that, under the frequently
met condition of a parabolic dependence of the energy on the
weak-moment magnitude, the weak moments and their interactions can be
eliminated via an analytical integration of the partition function in
favour of the strong moments with renormalized,
temperature-independent exchange constants, \new{with the
  renormalization accounting for the weak-moment fluctuations at
  non-zero temperatures.} We also show that the renormalized constants
are actually the ones probed by constrained spin-spiral calculations
of the strong-moment subsystem, thus simplifying
calculations. \new{Finally we show that the thermodynamic correlation
  functions of the full system including the strong and weak moments
  can be derived as polynomials of the correlation functions of the
  system of strong moments only but with renormalized interactions.}  This
renormalization will affect various quantities such as
temperature-dependent magnetization, susceptibility, or spin-stiffness
constant. \new{The method can prove useful for systems comprising $3d$
atoms with strong moments together with $4d$ or $5d$ atoms with weak
moments, such as transition-metal alloys, Heusler alloys or $3d$
overlayers deposited on $4d$ or $5d$ metal surfaces.}

\section*{Acknowledgments}

The authors would like to thank Leonid Sandratskii and Kyrill
Belashchenko for enlightening discussions. This work was supported
from funds by the EC Sixth Framework Programme as part of the European
Science Foundation EUROCORES Programme SONS under contract
N. ERAS-CT-2003-980409, and by the Young Investigators Group Programme
of the Helmholtz Association, Contract VH-NG-409. We gratefully
acknowledge sypport by the J\"ulich Supercomputing Centre.

\section*{Appendix}

We provide a derivation of a formula concerning the renormalization of
interactions $J^b_{ij}$ of the strong-moment subsystem, $\{\vc{M}_i\},
i\in\{1,\cdots,N_{\rm s}\}$ when it is in contact with a weak-moment
subsystem, $\{\vc{\mu}_l\}, l,l'\in \{1,\cdots,N_{\rm w}\}$, in the
presence of interactions $A_{ll'}$ among the moments of the weak
subsystem. The main complication in the presence of $A_{ll'}\neq 0$
for $l\neq l'$ is that the strong moments are interacting with a
system of coupled harmonic terms, instead of independent harmonic terms
which were treated in Sec.~\ref{Sec:longitudinal}. In particular we
adopt the following conventions. The Hamiltonian reads
\begin{equation}
\mathcal{H}=\mathcal{H}_{\rm s} + \mathcal{H}_{\rm w} +
\mathcal{H}_{\rm int}
\label{eq:A1}
\end{equation}
where the strong-system, weak-system and interacting parts are respectively:
\begin{eqnarray}
\mathcal{H}_{\rm s} &=& -\frac{1}{2}\sum_{\substack{ij\\ i\neq j}}J^b_{ij}
\vc{M}_i\cdot\vc{M}_j 
\label{eq:A2a}\\
\mathcal{H}_{\rm w} &=& \sum_{ll'}A_{ll'}
\vc{\mu}_l\cdot\vc{\mu}_{l'} 
\label{eq:A2b}\\
\mathcal{H}_{\rm int} &=& -\sum_{l} \sum_{n\in n(l)}
\frac{\kappa}{2a} \vc{M}_n\cdot \vc{\mu}_l
\label{eq:A2c}
\end{eqnarray}
In Eq.~(\ref{eq:A2b}), the diagonal part $A_{ll}=a>0$ is the quadratic
on-site energy term while the off-diagonal terms $A_{ll'}$ describe
intersite interactions between the weak moments.\cite{footnote3} In
case of ferromagnetic coupling it is expected that $A_{ll'}<0$ for
$l\neq l'$ (but with the determinant $\mathrm{det}|A_{ll'}|>0$).

The strategy is to eliminate the weak moments by integrating
analytically the weak plus intercting part of the partition function,
ending up with renormalized interactions of the strong moments. To
this end we take advantage of the identity
\begin{equation}
\begin{split}
\int dx_1\cdots dx_N
\exp\left[-\sum_{\lambda\lambda'}C_{\lambda\lambda'}x_{\lambda}x_{\lambda'}
  + \sum_{\lambda} b_{\lambda} x_{\lambda}\right] \\ 
=\frac{\pi^{N/2}}{\sqrt{\det C}} \exp\left[\sum_{\lambda\lambda'}(C^{-1})_{\lambda\lambda'}b_{\lambda} b_{\lambda'}\right]
\end{split}
\label{eq:A3}
\end{equation}
where $\det C$ is the determinant of the positive-definite matrix $C$.
It is convenient to use a combined index $\lambda=(l,\alpha)$ with
$\alpha\in \{xyz\}$ and define
$b_{\lambda}=\frac{1}{k_BT}\frac{\kappa}{2a}\sum_{n(l)}M_{l\alpha}$
and
$C_{\lambda\lambda'}=\frac{1}{k_BT}A_{ll'}\delta_{\alpha\alpha'}$. Then
$(C^{-1})_{\lambda\lambda'} = k_BT
(A^{-1})_{ll'}\delta_{\alpha\alpha'}$. Applying this to the partition
function $\mathcal{Z}$ yields
\begin{widetext}
\begin{eqnarray}
\mathcal{Z} &=& \int d\Omega_1\cdots d\Omega_{N_{\rm s}}
e^{-\mathcal{H}_{\rm s}/k_BT} \int d^3\mu_1\cdots d^3\mu_{N_{\rm w}}
e^{-(\mathcal{H}_{\rm w} + \mathcal{H}_{\rm int})/k_BT} 
\label{eq:A4.1}\\
&=&
\frac{(\pi k_BT)^{3N_{\rm w}/2}}{(\det A)^{3/2}}
\int d\Omega_1\cdots d\Omega_{N_{\rm s}} \exp\left[-\frac{1}{k_BT}\mathcal{H}_{\rm s}\right]
\exp\left[-\frac{1}{k_BT}\left(\frac{\kappa}{2a}\right)^2\sum_{ll'}(A^{-1})_{ll'}
  \sum_{n(l)}\sum_{n(l')}\vc{M}_n\cdot\vc{M}_{n'}\right] \\
\label{eq:A4.2}
&=&
\frac{(\pi k_BT)^{3N_{\rm w}/2}}{(\det A)^{3/2}}
\int d\Omega_1\cdots d\Omega_{N_{\rm s}}
\exp\left[-\frac{1}{k_BT}\mathcal{H}_r\right]
\label{eq:A5}
\end{eqnarray}
where a renormalized Hamiltonian of the strong-moment sublattice has been introduced, 
\begin{equation}
\mathcal{H}_r = -\frac{1}{2}\sum_{\substack{ij\\ i\neq j}}J^b_{ij}
\vc{M}_i\cdot\vc{M}_j + \left(\frac{\kappa}{2a}\right)^2\sum_{ll'}(A^{-1})_{ll'}
  \sum_{n(l)}\sum_{n'(l')}\vc{M}_n\cdot\vc{M}_{n'}.
\label{eq:A6}
\end{equation}
\end{widetext}
This reduces to $\mathcal{H}_r$ of Eq.~(\ref{eq:L10}) if $A_{ll'}=a\delta_{ll'}$.
Expression (\ref{eq:A6}) is obviously of the traditional Heisenberg type, but a
further reduction to a form with renormalized parameters,
$\mathcal{H}_r=-\frac{1}{2}\sum J^r_{ij}
\vc{M}_i\cdot\vc{M}_j$ requires knowledge of the specific geometry
of each problem taking into account the sums over neighbours of $l$
and $l'$, $\sum_{n(l)}$ and $\sum_{n'(l')}$. 

Thermal averages $\langle f(\{\vc{M}_i\})\rangle$ can be calculated by
\begin{equation}
\begin{split}
\langle f(\{\vc{M}_i\})\rangle = \frac{1}{\mathcal{Z}}\frac{(\pi
  k_BT)^{3N_{\rm w}/2}}{(\det A)^{3/2}}   
\\
\times \int d\Omega_1\cdots d\Omega_{N_{\rm
    s}}  f(\{\vc{M}_i\})
\exp\left[-\frac{1}{k_BT}\mathcal{H}^r_{\rm s}\right]
\end{split}
\label{eq:A7}
\end{equation}
thus the factor $\frac{(\pi k_BT)^{3N_{\rm w}/2}}{(\det A)^{3/2}}$ cancels and
the determinant $\det A$ need not be calculated. Thus one ends up with a usual
Heisenberg-model treatment. To gain some more insight, one can
recognize that in the presence of non-diagonal $A_{ll'}$ we have,
formally, a system of coupled harmonic oscillators interacting with
the strong moments. The normal modes of the coupled oscillators are
itinerant, and therefore the renormalized interactions are also
long-ranged. Even if $A_{ll'}$ are short ranged, making the matrix $A$
sparse, Eq.~(\ref{eq:A6}) shows that the renormalized interactions
involve the matrix $A^{-1}$, which is normally not sparse.

To complete the circle, it has to be shown for practical applications
that the renormalized interactions appearing in Eq.~(\ref{eq:A6}) are
the quantities that are probed by a DFT calculation where the
directions of the strong moments are constrained. In other words, in
such a DFT calculation the weak moments are allowed to relax to their
equilibrium values under the directional constraint on the strong
moments. The question is, if then the energy dependence (\ref{eq:A6})
is recovered, assuming that Eqs.~(\ref{eq:A2a}-\ref{eq:A2c}) are a
good approximation to the DFT energy landscape. The answer is
straightforward if we calculate the total energy of the constrained
Heisenberg Hamiltonian, i.e., without an integration over the
$\vc{M}_i$. The constrained partition function is just the last term
of Eq.~(\ref{eq:A4.1}), $\mathcal{Z}_1=\int d^3\mu_1\cdots
d^3\mu_{N_{\rm w}} e^{-(\mathcal{H}_{\rm w} + \mathcal{H}_{\rm
    int})/k_BT}$. The total energy at $T\rightarrow 0$ is
\begin{eqnarray}
E &=& \mathcal{H}_{\rm s}(\{\vc{M}_i\}) +\left.
\frac{kT^2}{\mathcal{Z}_1}\frac{\partial\mathcal{Z}_1}{\partial
  T}\right|_{T=0}\\
&=&  \mathcal{H}_{\rm s}(\{\vc{M}_i\}) \nonumber\\
&&+ \left(\frac{\kappa}{2a}\right)^2\sum_{ll'}(A^{-1})_{ll'}
  \sum_{n(l)}\sum_{n'(l')}\vc{M}_n\cdot\vc{M}_{n'}
\end{eqnarray}
{\it q.e.d.} This means that constrained (spin spiral) DFT
calculations on the strong sublattice are already corresponding to the
renormalized Hamiltonian and are therefore by virtue of
Eqs.~(\ref{eq:A4.1},\ref{eq:A5}) sufficient for the calculation of the
strong-sublattice thermal averages, without the need to calculate the
bare parameters or $A_{ll'}$.

However, if the weak-sublattice thermal averages are to be calculated,
one must additionally gain knowledge on the matrix $A_{ll'}$ of
Eq.~(\ref{eq:A2b}) as well as of the bare parameters $J^b$ of
Eq.~(\ref{eq:A2b}) and of $\kappa$ in Eq.~(\ref{eq:A2c}). The scheme
presented in Sec.~\ref{subsec:general} shows how this can be done in
the rather simple case of NiMnSb (where $A_{ll'}=a\delta_{ll'}$ is
diagonal), but in general this problem must be solved according to the
geometry and other factors in each case. In particular for the
calculation of the off-diagonal $A_{ll'}$, probably it is easiest to
calculate directly the susceptibility matrix $(A^{-1})_{ll'}$ by applying in
the DFT calculation a longitudinal external field on one atom and
probing the response in the moment of the neighboring atoms, or to
make a transformation to the normal modes in Fourier space and
calculate $A(\vc{q})$.

If all the ingredients are available, then one can calculate thermal
averages and correlation functions of the weak moments either by a
direct Monte-Carlo calculation, or by reducing these correlation
functions to correlations of the strong moments in the presence of
only the renormalized Hamiltonian, in an analogous way to the case of
Eq.~(\ref{eq:L18}), where the parameters $A_{ll'}$ and $\kappa$ must
be inevitably contained in the expansion coefficients. We present an
outline of how this is achieved in practice. The complication here
compared to Eq.~(\ref{eq:L18}) is that the matrix $A_{ll'}$ is not
diagonal. It is, however, real and symmetric, therefore one proceeds
by bringing it to a diagonal form $D$ with elements $D_l$. If $R$ is
the diagonalization matrix, then $D=RDR^{T}$, and we define the
transformation of the moment-coordinates $\tilde{\mu}_{q\alpha}=\sum_l
R_{ql}\mu_{l\alpha}$ and $\tilde{M}_{q\alpha}= \sum_n
R_{qn}M_{n\alpha}$. Also since $A_{ll'}$ is real and symmetric the
phase-space element is unchanged: $d^3\tilde{\mu}=d^3\mu$. (This
transformation is analogous to one bringing a system of coupled
harmonic oscillators in a normal-mode representation.)  Then one has,
for the arbitrary correlation function among the $xyz$-components of
the $N_{\rm w}$ weak moments,
$\langle(\mu_{1x})^{m_{1x}}\cdots(\mu_{N_{\rm w}z})^{m_{N_{\rm
      w}z}}\rangle$, the following integral:
\begin{widetext}
\begin{eqnarray}
\int d^3\mu_1 \cdots d^3\mu_{N_{\rm w}} 
(\mu_{1x})^{m_{1x}}\cdots(\mu_{N_{\rm w}z})^{m_{N_{\rm w}z}}
\exp\left\{\frac{1}{k_BT}\left[-\sum_{ll'\alpha}A_{ll'}\mu_{l\alpha}\mu_{l'\alpha} +
  \frac{\kappa}{2a}\sum_{l\alpha}\sum_{n(l)}
  M_{n\alpha}\mu_{l\alpha}\right]\right\} \\
=\int d^3\tilde{\mu}_1 \cdots d^3\tilde{\mu}_{N_{\rm w}}
(\sum_k R^{T}_{1k}\tilde{\mu}_{kx})^{m_{1x}}\cdots(\sum_k
R^{T}_{N_{\rm w}k}\tilde{\mu}_{k_wz})^{m_{N_{\rm w}z}} 
\exp\left\{\frac{1}{k_BT}\left[-\sum_{l\alpha}D_{l}\tilde{\mu}_{l\alpha}^2 +
  \frac{\kappa}{2a}\sum_{l\alpha}\sum_{n(l)}
  \tilde{M}_{n\alpha}\tilde{\mu}_{l\alpha}\right]\right\}  \\
=\prod_{l\alpha}\int d^3\tilde{\mu}_{l\alpha} (\sum_k
R^{T}_{lk}\tilde{\mu}_{k\alpha})^{m_{l\alpha}}
\exp\left\{\frac{1}{k_BT}\left[- D_{l}(\tilde{\mu}_{l\alpha} -
    \frac{1}{D_l}  \frac{\kappa}{2a}  \sum_{n(l)} \tilde{M}_{n\alpha}
    )^2 - \frac{1}{4D_l} ( \frac{\kappa}{2a}  \sum_{n(l)} \tilde{M}_{n\alpha})^2 \right]\right\}.
\end{eqnarray}
\end{widetext}
The last expression can be handled by expanding the term $(\sum_k
R^{T}_{lk}\tilde{\mu}_{k\alpha})^{m_{l\alpha}}$ and proceeding to an
analytic integration of powers of $\tilde{\mu}$ times the exponential
just as in Eq.~(\ref{eq:L18}). Mixed correlation functions between
strong and weak moments can also be reduced to strong-moment
correlation functions in the renormalized model via this prescribed
route.


\begin{thebibliography}{99}
\bibitem{RPA} Equations for the mean-field and random-phase
  approximations to the Curie temperature in multi-sublattice magnetic
  systems, including applications to Heusler alloys, are given, e.g.,
  in Ref.~\onlinecite{Sasioglu05}.
\bibitem{footnote} Due to a misprint in Ref.~\onlinecite{Lezaic06} the
last term of this equation has a spurious extra summation term. The
correct form is Eq.~(\ref{eq:L1}) here.
\bibitem{footnote1}
Sandratskii\cite{Sandratskii08}
  finds the same functional dependence of anergy and moment as in
  Eqs.~(\ref{eq:L5a},\ref{eq:L5b}), although the energy scale is
  smaller, perhaps due to the atomic sphere approximation.
\bibitem{footnote2} Unfortunately, this was not taken correctly in
Ref.~\onlinecite{Lezaic06}, resulting in a strong drop of the
calculated Ni moment and a peak of the Ni sublattice susceptibility at
50~K.
\bibitem{footnote3} Eq.~(\ref{eq:A2b}) is a special case of the
  Hamiltonian used by WGB\cite{Wysocki08} who additionally considered
  fourth-order terms for the study of longitudinal fluctuations. The
  WGB Hamiltonian is similar to the one used by Murata and
  Doniach,\cite{Murata72} but with the difference that Murata and
  Doniach consider a continuum model instead of a lattice model and a
  scalar instead of vector magnetization. Contrary to the continuum
  version of Murata and Doniach that shows ultraviolet
  divergencies,\cite{Schroter90} lattice Hamiltonians used by WGB and in
  Eq.~(\ref{eq:A2b}) have a natural wavelength cutoff, thus the
  thermal sums contain no divergencies.
\end{thebibliography}
\end{document}